\documentclass[
aps,%
12pt,%
final,%
notitlepage,%
oneside,%
onecolumn,%
nobibnotes,%
nofootinbib,%
superscriptaddress,%
noshowpacs,%
centertags]%
{revtex4}

\begin{document}
\selectlanguage{english}

\title{QUARK REGGE TRAJECTORY IN TWO LOOPS FROM UNITARITY RELATIONS}%

\author{\firstname{A.~V.}~\surname{Bogdan}}
\email{A.B.Bogdan@inp.nsk.su}

\author{\firstname{V.~S.}~\surname{Fadin}}
\email{Fadin@inp.nsk.su}
\affiliation{%
Institute for Nuclear Physics, 630090 Novosibirsk, Russia\\
and Novosibirsk State University, 630090 Novosibirsk, Russia
}%
\begin{abstract}
The two-loop quark Regge trajectory is obtained at arbitrary
space-time dimension $D$ using the $s$-channel unitarity
conditions. Although explicit calculations are performed for
massless quarks, the method used allows to find the trajectory for
massive quarks as well. At $D\rightarrow 4$ the trajectory turns
into one derived earlier from the high-energy limit of the
two-loop amplitude for the quark-gluon scattering. The comparison
of two expressions, obtained by quite different methods, serves as
a strict cross check of many intermediate results used in the
calculations, and their agreement gives a strong evidence of
accuracy of these results.
\end{abstract}
\maketitle
\section{Introduction}
Perturbative QCD is widely used  for the description of
semihard~\cite{semi-hard} as well as hard processes~\cite{hard}.
But whereas the theory of  the latter ones  is clear and plain, a
lot of problems remains unsolved for the former processes. The
applicability of perturbation theory, improved by the
renormalization group,  to a hard process with a large typical
virtuality $Q^2$ is justified by the smallness of the strong
coupling constant $\alpha_s(Q^2)$. In the semihard case, however,
smallness of the ratio $x$ of the typical virtuality $Q^2$ to the
squared c.m.s.\ energy  $s$ requires resummation of the terms
strengthened  by powers of $\ln(1/x)$. In the scattering channel
this problem is related to the theoretical description of high
energy amplitudes at fixed (not growing with $s$) momentum
transfer $t$. It turns out, that the Gribov-Regge theory of
complex angular momenta, which was developed much before
appearance of QCD, is eminently suitable for description of the
QCD amplitudes, due to the remarkable property of  QCD --
Reggeization of its elementary particles, gluons and
quarks~\cite{Grisaru}-\cite{FS}. The Reggeization means, in
particular, that with account of radiative corrections in the high
energy limit  $s$-dependence of QCD amplitudes with gluon ($G$) or
quark ($Q$) quantum numbers in the $t$--channel is given by Regge
factors $\left(s\right)^{j_P(t)}$, with $P=G$ or $P=Q$
accordingly. The functions ${j_P(t)}$ with the property
${j_P(m_P^2)=s_P}\,$  ($m_P$ and $s_P$ are respective mass and
spin values), called Regge trajectory,  describe motion  of poles
of corresponding $t$--channel partial waves in the complex angular
momentum plane. In this respect QCD  sharply differs from QED,
where only amplitudes with electron exchange in the
$t$-channel~\cite{electron}, but not with photon
one~\cite{photon}, acquire the Regge factors.

The Reggeization phenomenon is extremely important at high energy.
The gluon Reggeization is especially significant, since gluon
exchanges in the $t$-channel provide non-decreasing at large $s$
cross sections. In particular, the gluon Reggeization constitutes
the basis of the famous BFKL approach~\cite{FKL, BFKL} to the
theoretical description of high energy processes in QCD.
Formulated originally in leading logarithmic approximation (LLA),
the BFKL approach is developed now in next-to-leading one (for
references see, for instance,~\cite{F2002}),  since LLA is not
sufficiently reliable, especially because it does not fix scales
neither longitudinal (for $s$), nor transverse (for running
coupling $\alpha_s$) momenta. This development extensively uses
the gluon Reggeization, which has been proved in LLA~\cite{BLF},
but in next-to-leading approximation (NLA) till now remains  a
hypothesis, although successfully  passed through a set of
stringent tests on self-consistency (see, for
instance,~\cite{F2003} and references therein).  Accordingly, the
next-to-leading order (NLO) gluon Regge trajectory and Reggeized
gluon vertices are calculated; moreover, a way  for the proof of
the gluon Reggeization in  NLA is outlined (see, e.g.,
~\cite{F2002} and references therein).

Along with the Pomeron, which appears in the BFKL approach as a
compound state of two  Reggeized gluons, the hadron phenomenology
requires Reggeons, which can be constructed in QCD as  colorless
states of  Reggeized quarks and antiquarks. It demands further
development of the  Reggeized quark theory which remains in a
worse state than  the Reggeized gluon theory, although a
noticeable progress was achieved last years. In particular,
multi-particle Reggeon vertices required in NLA were found
\cite{LV} and the NLO corrections to the LLA vertices were
calculated \cite{FF01, KLPV} assuming the quark Reggeization in
NLA. Note that the Reggeization hypothesis is extremely powerful;
but in the quark case actually it was not proved even in LLA,
where merely its self-consistency was shown, in all orders of
$\alpha_s$ but only in a particular case of elastic quark-gluon
scattering~\cite{FS}. Recently  the hypothesis was tested at NLO
in order $\alpha_s^2$ in~\cite{BD-DFG}, where its compatibility
with the two-loop amplitude for the quark-gluon scattering was
exhibited and the NLO correction to the quark trajectory was found
in the limit of the space-time dimension $D$ tending to the
physical value $D=4$.

In this paper we  investigate  the quark Reggeization in NLO by
the  method  based on the $s$-channel unitarity and the
analyticity of scattering amplitudes, which was developed  for
analysis of processes with gluon exchanges~\cite{Lipatov76, FKL}
and was already successfully applied to processes with fermion
exchanges~\cite{FS}. The two-loop quark trajectory at arbitrary
space-time dimension $D$ is obtained as a particular result of the
investigation. At $D\rightarrow 4$ the trajectory goes into one
derived in~\cite{BD-DFG}.  This agreement testifies to accuracy of
many intermediate results used in both derivations. In the method
used here the trajectory is obtained from the requirement of the
compatibility of the Reggeized form of the amplitudes with the
$s$-channel unitarity at the two-loop level. A possible
generalization of this requirement to all orders of perturbation
theory should give the "bootstrap" conditions on the Reggeized
quark vertices and the trajectory in QCD. Verification of them
will give a strict test for the quark Reggeization.  A proof of
the Reggeization is also possible on this way.

The calculation of the two-loop corrections to the quark
trajectory is performed explicitly for massless quarks; but  the
method used here allows to do it for massive quarks as well, since
all necessary one--loop Reggeon vertices for the massive case are
known now.

The paper is organized as follows. In the next Section all
necessary denotations are introduced and the method of calculation
is discussed. Section 3 is devoted to the calculation of the
two--particle contribution to the $s$--channel discontinuity of
the quark--gluon scattering amplitude. The contribution of the
three--particle intermediate state is calculated in  Section 4.
The final expressions for the discontinuity and the two-loop
corrections to the quark trajectory are presented and discussed in
Section 5. For convenience, the integrals encountered  in Sections
3 and 4 are listed in Appendices ~\textbf{A} and \textbf{B}
respectively. Details of the calculation of a new nontrivial
integral arising in present calculations are given in Appendix
\textbf{C}.
\section{Denotations and method of calculation}
\begin{figure}[t!]
\setcaptionmargin{0mm}
\onelinecaptionsfalse
  \includegraphics[width=40mm]{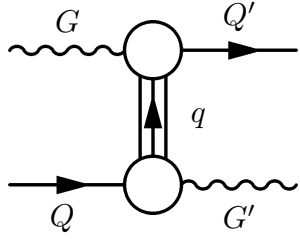}
\captionstyle{flushleft}\caption{Schematic representation of the backward quark--gluon
scattering process $G+Q\rightarrow Q'+G'$. The triple line denotes
an intermediate $t$--channel state with momentum
$q=p_{Q}-p_{G'}=p_{Q'}-p_G$.}
  \label{pic1}
\end{figure}

Let us consider the backward quark-gluon scattering process (see
Fig.~\ref{pic1}) in the limit of large (tending to infinity)
c.m.s. energy and fixed momentum transfer $t\equiv
q^2=(p_Q-p_{G'})^2$.

We  use the Sudakov decomposition of momenta
$$
p=\beta  p_1+\alpha p_2 +p_{\perp }~,
$$
\begin{equation}
p_1^2=p_2^2=0~,\;\;(p_1+p_2)^2 =s~, \,\,\,s\alpha \beta =p^2
-p_{\perp }^2~, \label{sud}
\end{equation}
supposing that the momenta $p_G,\;p_{Q'}$ and $p_Q,\;p_{G'}$ are
close to the light-cone momenta $p_1$ and $p_2$ respectively, that
is
\begin{equation}
\beta_G\sim\beta_{Q'}\sim\alpha_Q\sim\alpha_{G'} \simeq 1,\;\;
\beta_Q\simeq\beta_{G'}\simeq\alpha_G\simeq\alpha_{Q'}
\sim\frac{|t|}{s},
\end{equation}
and all transverse momenta are limited, so that $q\simeq
q_{\perp}$. We don't suppose that $p_1$ and $p_2$ are contained in
the initial momentum plane, in order to maintain symmetry between
cross channels  and to make more evident substitutions for
transitions between channels. For a gluon having momentum $k_a$
($k_b$) with predominant component along $p_1$ ($p_2$) we use
physical polarization vectors in the light-cone gauge
$e(k_a)p_2=0$ ($e(k_b)p_1=0$), so that
\begin{equation}
e(k_a)=e(k_a)_{\perp}-\frac{(e(k_a)k_{a})_{\perp}}{k_ap_2}p_2~,\;\;
e(k_b)=e(k_b)_{\perp}-\frac{(e(k_b)k_{b})_{\perp}}{k_bp_1}p_1~,
\label{axial2}
\end{equation}
where  $(a b)_{\perp}$ means $(a_{\perp} b_{\perp})$. Furthermore,
since with our choice of gauges gluon polarization vectors are
expressed in terms of their transverse components, from now we
will use only these components, without explicit indications, so
that everywhere below $e$ means $e_{\perp}$. The same we will do
for the momentum transfer $q$.

The large $s$ and fixed $t$ limit of scattering amplitudes is
related to quantum numbers in the $t$--channel. For the gauge
group $SU(N_c)$   $t$-channel colour state of the process depicted
at Fig.~\ref{pic1} contains three irreducible representations of
the colour group (for QCD with $N_c=3$ it is
$\underline{3}\oplus\underline{\bar 6}\oplus\underline{15}$).
Therefore it is natural to decompose the quark-gluon scattering
amplitude into three parts, in accordance with the
representations~\cite{FS, BD-DFG}. At the same time, in the
Gribov--Regge theory each part must be decomposed into two pieces
according to the new quantum number -- signature, which is
introduced in the theory. Besides quantum numbers, commonly used
for particle classification, a Reggeon has  definite signature,
positive or negative, which is actually a parity of the
$t$--channel partial waves  in respect of the substitution
$\cos\theta_t\rightarrow -\cos\theta_t $ (that turns into
$s\rightarrow u = -s$ in the limit of large $s$ and fixed $t$).
Consequently, there are six terms $A^{(\pm)}_\chi~,
\;\;\chi=\underline{3}~,\underline{\bar 6}$ and $\underline{15}$
in full decomposition~\cite{FS, BD-DFG} of the quark-gluon
scattering amplitude. As it is known~\cite{FS}, in LLA only one of
the positive signature amplitudes, namely
$A^{(+)}_{\underline{3}}$, does survive;  and at the same time it
has the Reggeized form. It is not so for the negative signature
amplitudes. The Bethe-Salpiter type equation obtained for them in
LLA~\cite{FS} has not a simple Regge-type solution (in fact, no
solution has been found at all).  Note that they are not actually
leading in each order of perturbation theory, because leading
logarithms cancel in them as the result of antisymmetry with
respect to the exchange $s\rightarrow u = -s$. Below we consider
only the amplitudes of positive signature. As we will see, in NLA,
as well as in LLA, only  amplitudes with a colour triplet in the
$t$--channel survives out of them. For the quark-gluon scattering
the contribution of the Reggeized quark  can be represented as
(here and below we write symbols of initial (final) particles as
lower (upper) indices of scattering amplitudes and put on the
first place particles with momenta close to $p_1$):
\begin{equation}\label{q_rege}
{\cal R}^{Q'G'}_{G\,Q}=\Gamma_{Q'G}
\frac{1}{m_Q-\hat{q}}%
\frac{1}{2}\left[\left(\frac{-s}{-t}\right)^{\delta_{\mathrm{T}}(\hat{q})}+%
\left(\frac{s}{-t}\right)^{\delta_{\mathrm{T}}(\hat{q})}\right]\Gamma_{G'Q}\,,
\end{equation}
where $\Gamma_{Q'G}$ and $\Gamma_{G'Q}$ are effective vertices for
interaction of particles (quarks and gluons) with the Regeized
quark; we call them  PPR vertices, and we call
$\delta_{\mathrm{T}}(\hat{q})$ --- the quark trajectory. Strictly
speaking, for massive quarks there are two trajectories, in
accordance with two possible parity states for a off mass shell
quark. These trajectories are determined  by eigenvalues of
$\delta_{\mathrm{T}}(\hat{q})$. We perform actual calculations for
the massless case, when $\delta_{\mathrm{T}}(\hat{q})$ depends in
fact not  on $\hat{q}$, but on $\hat{q}^2=t$, and  write it as
$\delta_{\mathrm{T}}(t)$. Note, however, that the PPR vertices
$\Gamma_{Q'G}$ and  $\Gamma_{G'Q}$ are known now for massive
quarks~\cite{KLPV}, so that all consideration presented below can
be transferred on the massive case in a straightforward way.

We  demonstrate  that the Reggeized form (\ref{q_rege}) is
compatible with the $s$-channel unitarity and  obtain the NLO
contribution to the trajectory $\delta_{\mathrm{T}}$. More
precisely, we  calculate, using the unitarity relation,  both
logarithmic and non-logarithmic terms in the  two-loop
$s$--channel discontinuity of the backward quark-gluon scattering
amplitude with positive signature and prove that only  colour
triplet $t$--channel states contribute to the discontinuity. It
means that only the colour triplet part $A^{(+)}_{\underline{3}}$
of the amplitude does survive at NLO as well as at LO. We compare
the calculated discontinuity with the discontinuity of the
Reggeized quark contribution (\ref{q_rege}). The logarithmic terms
of confronted discontinuities turn out  to be equal. The
non-logarithmic terms in the discontinuity of (\ref{q_rege}) are
expressed through the one--loop corrections to the PPR vertices,
which are known, and the two-loop correction to the trajectory,
that makes possible to obtain the last correction.

For  massless quarks the PPR vertices entering in (\ref{q_rege})
have the form~\cite{FF01}:
\begin{align}
\Gamma_{Q'G}&=-g\bar{u}(p_{Q'})t^G\left[\hat{e}_{G}(1+\delta_e(t))+\frac{{(e_G q)}\hat{q}}%
{q^2}\delta_q(t)\right]\,,\nonumber\\
\Gamma_{G'Q}&=-g\left[\hat{e}^\ast_{G'}(1+\delta_e(t))+\frac{{(e^\ast_{G'}q)}\hat{q}}%
{q^2}\delta_q(t)\right]t^{G'}u(p_Q)\,.\label{q vertex}
\end{align}
Here $t^G$, $t^{G'}$ are the color group generators in the
fundamental representation; we omit colour wave functions for
gluons and assume that they are included in
$u(p_Q),\;\bar{u}(p_{Q'})$ for quarks. The one--loop corrections
$\delta_e(t)$, $\delta_q(t)$  can be written as
\begin{equation}\label{corr}
\delta_e(t)=\om\,\delta_e^{(1)}\,,\quad
\delta_q(t)=\om\,\delta_q^{(1)}\,,
\end{equation}
where
\begin{gather}\label{q vertex_corr}
\delta_e^{(1)}=\frac{C_F}{2N_c}\left(\frac{1}{\epsilon}-\frac{3(1-\epsilon)}{2(1+2\epsilon)}+\psi(1)+\psi(1-\epsilon)-2\psi(1+\epsilon)\right)+%
\frac{1}{2\epsilon}-\frac{\epsilon}{2(1+2\epsilon)}\,,\\%
\delta_q^{(1)}=\frac{\epsilon}{2(1+2\epsilon)}\left(1+\frac{1}{N_c^2}\right)\nonumber\,,
\end{gather}
$\psi(x)=\Gamma'(x)/\Gamma(x)$ is the logarithmic derivative of
the Euler gamma function, and $\om$ is the one--loop gluon Regge
trajectory:
\begin{equation}\label{omega1}
\quad\om=%
-\frac{2N_c}{\epsilon}\frac{g_s^2}{(4\pi)^{2+\epsilon}}\left(\frac{\mu^2}{-t}\right)^{-\epsilon}
\!\!\Gamma_{\epsilon}\,,\quad\Gamma_{\epsilon}=
\frac{\Gamma^2(1+\epsilon)\Gamma(1-\epsilon)}{\Gamma(1+2\epsilon)}\,.
\end{equation}
We use conventional dimensional regularization; the space-time
dimension $D=4+2\epsilon$ and $g_s=g\mu^{\epsilon}$ is the
dimensionless bare coupling constant.

To avoid uncertainties let us note that the vertices
$\Gamma_{G\bar Q}$ and $\Gamma_{\bar Q G'}$ are obtained from
(\ref{q vertex})  by the substitutions $\bar u(p_{Q'})\rightarrow
\bar v(p_{\bar Q})~, \;\;e_G\rightarrow e^*_G\;\;$ and
$\;\;u(p_{Q})\rightarrow v(p_{\bar Q})~, \;\;e^*_{G'}\rightarrow
e_{G'}$ respectively.

Representing the quark trajectory  as
\begin{gather}
\delta_{\mathrm{T}}(t)=\frac{\om}{N_c}\,\dt{1}+%
\frac{1}{2}\left(\frac{\om}{N_c}\right)^2\dt{2}\,, \label{g-traj}
\end{gather}
where $\dt{1}=C_F$ \cite{FS}, and  the two--loop $s$--channel
discontinuity of the Reggeized quark contribution (\ref{q_rege})
as
\begin{equation}\label{disc}
\left[{\cal R}^{Q'\,G'}_{G\,Q}
\text{(two--loop)}\right]_s%
=\frac{i\pi
g^2}{t}\frac{1}{2}\left(\frac{\om}{N_c}\right)^2\bar{u}(p_{Q'})t^Gt^{G'}\,\Delta_R\,u(p_Q)\,,
\end{equation}
we have from \eqref{q_rege}, \eqref{q vertex} \eqref{g-traj} and
\eqref{corr} \begin{equation}
\label{Delta}
\Delta_R=\dt{2}\Qc+2C_F\left(2N_c\delta_e^{(1)}+C_F\Ln\right)\Qc+%
2C_FN_c\delta_q^{(1)}\Ec\,,
\end{equation}
where
\begin{equation}\label{QE-def}
\Qc=\hat{e}_{G}\hat{q}\hat{e}^\ast_{G'}\,,\quad
\Ec=\hat{e}_{G}{(e^\ast_{G'}q)}+\hat{e}^\ast_{G'}{(e_G q)}\,,
\end{equation}
and $\delta_e^{(1)}$, $\delta_q^{(1)}$ are given by (\ref{q
vertex_corr}).

Below we calculate the discontinuity
$\left[\left(\Ac^{(+)}\right)^{Q'\,G'}_{G\,Q}\text{(two--loop)}
\right]_s$ of the backward quark-gluon scattering amplitude with
 positive signature from the $s$--channel unitarity condition.
We show, that it has the same colour structure as $\left[{\cal
R}^{Q'\,G'}_{G\,Q} \text{(two--loop)}\right]_s$ (\ref{disc}), that
is in this approximation only the colour triplet state survives in
the positive signature.  Writing the calculated discontinuity in
the same form as the right hand side of (\ref{disc}) with
$\Delta_s$ instead of $\Delta_R$, we see that $\Delta_s$ has the
same helicity structure as $\Delta_R$ (\ref{Delta}) and that their
logarithmic terms coincide; moreover, the non-logarithmic terms at
the helicity non-conserving structure $\Ec$ also do coincide.
After that the requirement of equality of the non-logarithmic
terms at the helicity conserving structure $\Qc$ gives us
$\dt{2}$.

The calculation of $\Delta_s$ is the main content of this paper.
It is determined from the $s$--channel unitarity relation:
\begin{align}
\left[\left(\Ac^{(+)}\right)^{Q'\,G'}_{G\,Q}\text{(two--loop)}
\right]_s&=\frac{i\pi
g^2}{t}\frac{1}{2}\left(\frac{\om}{N_c}\right)^2\bar{u}(p_{Q'})t^Gt^{G'}
\,\Delta_s\,u(p_Q)\\
\mbox{}&=i\mathrm{P}^{(+)}\sum\limits_n \int\!\! d\Phi_n \Ac^n_{G\, Q}%
\left(\Ac^n_{Q'\,G'}\right)^\ast\,,\label{Disc}
\end{align}
where $\Phi_n$ is the  $n$--particle phase space element and
$\mathrm{P}^{(+)}$ is the positive signature projector. The
summation is performed over  two-- and three--particle
intermediate states; accordingly,  we represent the discontinuity
as the sum of two contributions
\begin{equation}\label{Delta_sum_delta}
\Delta_s=\Delta_s^{(2)}+\Delta_s^{(3)}\,.
\end{equation}
The projection on positive signature means  the half--sum of the
$s$--channel discontinuities for the direct ($G Q\rightarrow
Q'G'$) and the crossed ($ G \tilde G\rightarrow  Q'\bar Q'$)
processes. More precisely, if one represents the discontinuity of
the direct processes as $\langle Q'G'|M|G Q\rangle$, where $|G Q\rangle$ is a
spin and colour quark--gluon wave function, and the discontinuity
for the process $G \tilde G\rightarrow  Q'\bar Q'$ with $p_{\tilde
G}=p_Q,\;p_{\bar Q'}=p_{G'}$ as $\langle Q'\bar Q' |M_c|G\tilde G\rangle$,
then the projection on  positive signature is $\langle
Q'G'|(M+M_c)/2|GQ\rangle$. Note that calculating $\Delta_s$ we always
use that rightmost $\hat p_2$ and leftmost $\hat p_1$ in $M$  and
$M_c$ give negligible contributions  because of the Dirac
equation.
\section{Two--particle contribution to the discontinuity}
In the direct channel a two--particle intermediate state can be
solely quark--gluon one. Since only limited transverse momenta of
intermediate particles are important in the unitarity relation (we
will see it directly; actually, it is a consequence of the
renormalizability), non--negligible contributions are given by two
non--overlapping kinematical regions.  In one of them intermediate
gluon momentum  is close to $p_G$ (see Fig. 2a),  and in another
to $p_Q$ (Fig. 2b). In both cases the amplitudes in the right hand
side of the unitarity relation (\ref{Disc}) are in  Regge type
kinematics.

As we need to calculate the two--loop contribution to the
discontinuity, one of  them has  to be taken in the Born
approximation and another one  in the one--loop approximation.  An
important point is that since Born amplitudes are real, only real
parts of one--loop amplitudes are essential for the calculation of
the discontinuity. Therefore required amplitudes are determined by
Reggeized quark and gluon contributions. Moreover, we can use
$\Ac_n^{Q'G'}$ instead of $\left(\Ac^n_{Q'G'}\right)^\ast$, as
imaginary parts of the amplitudes are not important.

The amplitudes with $t$--channel quarks can be obtained by evident
substitutions from \eqref{q_rege}. Using (\ref{q vertex}),
(\ref{corr}) for the process $G+Q\rightarrow Q'+G'$  we have with
required accuracy:
\begin{multline}
A_{G\,Q}^{Q'G'}= \frac{g^2}{-q^2} \bar{u}(p_{Q'})t^G
t^{G'}\left\{\hat{e}_{G}\hat{q}\hat{e}^\ast_{G'}
\left[1+\omega^{(1)}(q^2)\left(2\delta_e^{(1)}+%
\frac{C_F}{N_c}\ln\frac{s}{-q^2}\right)\right]
\right.\\
\mbox{}+\left.\omega^{(1)}(q^2)\delta_q^{(1)}\left[\hat{e}_{G}
{(e^\ast_{G'}q)}+{(e_Gq)}\hat{e}^\ast_{G'}\right]
\right\}u(p_{Q})~, \label{q-oneloop}
\end{multline}
where $q=(p_Q-p_{G'})_{}$.

In the following in the amplitudes at the right hand side of
(\ref{Disc}) we denote Reggeized gluon and quark momenta $q_1$ and
$q_2$ respectively, $q_1+q_2=q$, and Reggeized gluon colour index
$r$. Since $q_1\simeq q_{1{\perp}}~, \;\;q_2\simeq q_{2\perp}~,$
everywhere below we omit the sign $\perp$ at $q_{1,2}$, so that
$q_{i}$ means $q_{i\perp}~, \;\; i=1,2$.

Amplitudes for  processes $AB\rightarrow A'B'$ with  gluon
exchanges  with our accuracy are written as
\begin{equation}
{\cal A}_{A\,B}^{A'B'} = \frac{2s}{q^2_{1}}\Gamma _{{A'}A}^{r}
\left(1+\omega^{(1)} (q^2_{1})\ln\frac{s}{-q^2_{1}} \right)\Gamma
_{{B'}B}^{r}~, \label{g-regge}
\end{equation}
where $\Gamma _{{A'}A}^{r}$ and $\Gamma _{{B'}B}^{r}$ are
Reggeized gluon vertices, which can be found in \cite{F2003}. In
the direct channel we need quark--gluon scattering amplitudes. For
the process $G+Q\rightarrow G_1+Q_1$ (see the left part of Fig.
2a) we obtain
\begin{multline}
A_{G\,Q}^{G_1Q_1}= \frac{2g^2s}{q^2_{1}} \T{r}{G_1}{G}\bar
u(p_{Q_1})t^r\frac{\hat p_1}{s}u(p_Q)\\
\mbox{}\times\left\{-(e^\ast_{G_1}e_G)\left[1+\omega^{(1)}(q^2_{1})\left(\delta^{(1+)}_G+\delta^{(1-)}_G+
\delta^{(1)}_Q+\ln\frac{s}{-q^2_{1}}\right)\right]\right.\\
\mbox{}+\left.\omega^{(1)}(q^2_{1})(D-2)\frac{{(e^\ast_{G_1}q_1)}{(e_Gq_1)}
}{q^2_{1}}\delta^{(1-)}_G\right\}\,, \label{a-g}
\end{multline}
where $\T{r}{G_1}{G}$ are colour generators in the adjoint
representation, $q_1=(p_{G_1}-p_G)_{\perp}$,$\;\; \delta_Q$
represents the one loop corrections to the Quark--Quark--Reggeon
vertex,
\begin{multline}\label{delta_Q}
\delta_{Q}^{(1)}=\frac{1}{2}\left[\frac{1}{\epsilon}+\psi(1-\epsilon)+\psi(1)-%
2\psi(1+\epsilon)+\frac{2+\epsilon}{2(1+2\epsilon)(3+2\epsilon)}\right.\\
\mbox{}-\left.\frac{1}{2N_c^2}\left(1+\frac{2}{\epsilon(1+2\epsilon)}\right)-%
\frac{n_f}{N_c}\frac{1+\epsilon}{(1+2\epsilon)(3+2\epsilon)}\right]\,,%
\end{multline}
$n_f$ is the number of quark flavours;  $\delta_{G}^{(1+)}$ and
$\delta_{G}^{(1+)}$ represent helicity conserving and helicity
violating corrections to the Gluon--Gluon--Reggeon vertex;
\begin{multline*}
\delta_{G}^{(1+)}=\frac{1}{2}\left[\frac{2}{\epsilon}+\psi(1-\epsilon)
+\psi(1)-2\psi(1+\epsilon)\right.\\
\mbox{}-\left.\frac{9(1+\epsilon)^2+2}{2(1+\epsilon)(1+2\epsilon)(3+2\epsilon)}
+\frac{n_f}{N_c}\frac{(1+\epsilon)^3+\epsilon^2}{(1+\epsilon)^2(1+2\epsilon)(3+2\epsilon)}\right]\,,
\end{multline*}
\begin{equation}\label{delta_G}
\delta_{G}^{(1-)}=\frac{\epsilon}{2(1+\epsilon)(1+2\epsilon)(3+2\epsilon)}\left(-1+\frac{n_f}{N_c(1+\epsilon)}\right)\,.
\end{equation}
Note, that in order  to obtain $A^{Q'G'}_{Q_1G_1}$ (see the
right part of Fig. 2b) from (\ref{a-g}) one has to change $\hat
p_1$ to $\hat p_2$ besides evident substitution of symbols .

In addition to presented amplitudes, in the crossed channel we
need gluon--gluon and quark--antiquark forward scattering amplitudes
(see Fig. 3).

The first (see the left part of Fig. 3a) is written as
\begin{multline}
A_{G\,\tilde G}^{G_1 \tilde G_1} = \frac{2g^2s}{q^2_{1}}
\T{r}{G_1}{G}\T{r}{\tilde G_1}{\tilde G} \left\{(e^\ast_{G_1}e_G)
(e^\ast_{\tilde G_1}e_{\tilde G})
\left[1+\omega^{(1)}(q^2_{1})\left(2\delta^{(1+)}_G
+2\delta^{(1-)}_G+
\ln\frac{s}{-q^2_{1}}\right)\right]\right.\\
\mbox{}-\left.\omega^{(1)}(q^2_{1})(D-2) \frac{\left((e^\ast_{G_1}e_{G})
(e^\ast_{\tilde G_1}q_1){(e_{\tilde
G}q_1)}+(e^\ast_{G_1}q_1){(e_Gq_1)}(e^\ast_{\tilde G_1}e_{\tilde
G}) \right)}{q^2_{1}}\delta^{(1-)}_G\right\}\,, \label{a-gg}
\end{multline}
and the second (see the right part of Fig. 3b)
\begin{equation}
A_{Q_1\bar{Q}_1}^{Q'\bar Q'}= -\frac{2g^2s}{q^2_{1}} \bar
u(p_{Q'})t^r\frac{\hat p_2}{s}u(p_{Q_1})\bar v(p_{\bar{Q}_1})t^r\frac{\hat
p_1}{s}v(p_{\bar Q'})
\left[1+\omega^{(1)}(q^2_{1})\left(2\delta^{(1)}_Q+\ln\frac{s} {-
q^2_{1}} \right)\right] \,. \label{a-qq}
\end{equation}

Before to start with calculation let us show that only a colour
triplet state survives in the discontinuity, due to cancellation
of contributions of all other colour states in the direct and
crossed channels. In fact, this cancellation has the same nature
as in leading order \cite{FS}. The matter is that if a one--loop
contribution is taken for one of the PPR vertices in Figs.~2 and 3,
then all other vertices must be taken at NLO  in Born
approximation. Therefore either both lower, or  both upper
vertices are Born ones.  Let us consider the first case. Since
upper parts of the diagrams Fig. 2a and Fig. 3a are equal,
contributions to $M+M_c$ from the lower lines enter as the sum
\begin{equation}
\gamma^{\mu}_{\perp}t^{G'}\sum_{\lambda}u^{\lambda}(p_{Q_1}) \bar
u^{\lambda}(p_{Q_1})t^r\frac{\hat p_1}{s} - \sum_{\lambda} \hat
e^\lambda_{\tilde G_1} t^{\tilde G_1}\left(e^{\lambda}_{\tilde
G_1} \right)^{\star\,\mu} T^r_{\tilde G_1 G'} =
\gamma^{\mu}_{\perp}t^rt^{G'}.
\end{equation}
Here we have omitted terms with leftmost $\hat p_1$ because of the
reason explained above,  and have taken the same Lorentz and
colour indices of the gluons $G'$ and $\tilde G$ as we don't write
their wave functions. It's easy to see  that lower lines of Fig.~2b
and Fig.~3b give in sum the same result. Since $t^r t^{G'}$
projects the $t$--channel quark--gluon state on a colour triplet, it
means that contributions of another colour states cancel in the
sum of the direct and cross channels. The case when  both upper
vertices are Born ones can be considered quite analogously.  One
can come to the same conclusion seeing that  sum of contributions
of Fig.~2a and Fig.~3b, as well as Fig.~2b and Fig.~3a,  is
proportional $t^{G}t^r$. It is not difficult to understand that
the cancellation of colour states different from triplet is not
restricted by considered diagrams and by the two--loop
approximation, but is a general property of NLA, as well as LLA.

Since we have shown that only a colour triplet survives in the
$t$--channel we can write
\begin{equation}\label{ampl_2}
\mathrm{P}^{(+)}_3\sum\limits_{P_1P_2}\Ac^{P_1P_2}_{G\,Q}
\left(\Ac^{P_1P_2}_{Q'G'}\right)^\ast%
=g^4\:\om\,s\,\bar{u}(p_{Q'})t^Gt^{G'}\Mc^{(2)}u(p_Q)\,,
\end{equation}
With the amplitudes listed above calculation of $\Mc^{(2)}$ is
straightforward. Dividing it into two pieces,
$\Mc^{(2)}=\Mc^{(2)}_\text{$Q$}+\Mc^{(2)}_\text{$G$}$, one of
which  contains one--loop corrections for a  quark channel, and
another for a gluon channel, we obtain
\begin{equation}
\Mc^{(2)}_\text{$Q$}=\left(\frac{q^2}{q^2_{2}}
\right)^{-\epsilon}\!\!\!\!%
\frac{C_F}{q_{1}^{2}q^{2}_{2}}\left\{\left(2\delta_e^{(1)}+%
\frac{C_F}{N_c}\ln\frac{s}{-q^2_{2}}\right)%
\hat{e}_{G}\hat{q}_{2}\hat{e}^\ast_{G'}+\delta_q^{(1)}[\hat{e}_G{(e^\ast_{G'}q_2)}+%
{(e_G q_2)}\hat{e}^\ast_{G'}]\right\}\,,
\end{equation}
and
\begin{multline}
\Mc^{(2)}_\text{$G$}=\left(\frac{q^2}{q^{2}_{1}}
\right)^{-\epsilon}\!\!\!\!\frac{1}{q_{2}^{2}%
q^{2}_{1}}\left\{\left[-\frac{1}{N_c}\delta^{(1)}_Q+N_c\left(
\delta_G^{(1+)}+%
\delta_G^{(1-)}\right)+C_F\ln\frac{s}{-q^2_{1}}\right]%
\hat{e}_{G}\hat{q}_{2}\hat{e}^\ast_{G'}\right.\\
\mbox{}-\left.\frac{N_c}{2}\frac{D-2}{q^2_{1}}\delta^{(1-)}_{G}%
[\hat{e}_{G}\hat{q}_{2}\hat{q}_{1}{(e^\ast_{G'}q_1)}+%
{(e_G q_1)}\hat{q}_{1}\hat{q}_{2}\hat{e}^\ast_{G'}]\right\}\,.%
\end{multline}
Integration over the phase space element
$d\Phi_2={d^{D-2}q_{2}}/({2s(2\pi)^{D-2}})$ is quite simple. The
integrals are well convergent at large $|q_{2}|$, so that the
integration region can be expanded to infinity. For convenience of
a reader we present necessary formulas in Appendix \textbf{A}. As
a result, we obtain for the two-particle contribution to
$\Delta_s$
\begin{multline}\label{Delta 2}
\Delta^{(2)}_s=\frac{2N_c\epsilon(-q^2)^{1-\epsilon}}{\Gamma_{\epsilon}\pi^{(D-2)/2}}\;%
\int\!\! d^{D-2}q_{2}\:\Mc^{(2)}=
2N_c\Xr\left\{2C_F\delta_e^{(1)}+\frac{N_c}{2}%
\left(\delta^{(1+)}_G-\frac{2\epsilon}{1-\epsilon}\delta_G^{(1-)}\right)\right.\\
\mbox{}-\left.\frac{1}{2N_c}\delta_Q^{(1)}
+\frac{C_F^2}{N_c}\left(\Ln+\Psi_1\right)+%
\frac{C_F}{2}\left(\Ln+\Psi_1+\frac{1}{2\epsilon}\right)\right\}\Qc\\
\mbox{}+\Xr\left\{C_F\delta_q^{(1)}+N_c\,\frac{\,\epsilon(1+\epsilon)}{1-\epsilon}\,
\delta_G^{(1-)}\right\}\Ec\,,
\end{multline}
where%
\begin{equation}\label{Xr}
\Xr=\frac{\Gamma(1-2\epsilon)\Gamma^2(1+2\epsilon)}{\Gamma(1+\epsilon)\Gamma(1+3\epsilon)\Gamma^2(1-\epsilon)}\,,
\end{equation}
and%
\begin{equation}
\Psi_1=\psi(1-2\epsilon)+\psi(1+3\epsilon)-\psi(1+2\epsilon)-\psi(1-\epsilon)\,.%
\end{equation}
\section{Tree--particle contribution to the discontinuity}
Let us denote  intermediate particles in the unitarity condition
(\ref{Disc}) $P_j$ and their momenta $k_j,\; j=1\div3$. Just as
before, only limited transverse momenta are important. As for
longitudinal ones, let us put (without loss of generality)
$\alpha_3\simeq 1$, that is for the direct process $G+Q\rightarrow
Q'+G'$ a particle $P_3$ is produced in the fragmentation region of
the initial quark (note that for the crossed process $ G+\tilde
G\rightarrow Q'+\bar Q'$ it is the region of $\tilde G$
fragmentation). Then $\beta_1+\beta_2\simeq 1$, i.e. at least one
of particles $P_i,\; i=1,2,\;$ is produced in the gluon
fragmentation region. Let it will be $P_1$; then $\beta_1 \sim 1$,
but $\beta_2$ can vary from $\beta_2 \sim 1$ (that means $P_2$ as
well as $P_1$ is produced in the gluon fragmentation region) to
$\beta_2 \sim |k^2_{2{\perp}}|/{s}$ (it means $\alpha_2\sim 1$,
i.e. $P_2$ is in the quark fragmentation region). Of course, the
same can be said with substitution $1\leftrightarrow 2$. Note that
region $1 \gg\beta_i \gg |k^2_{i{\perp}}|/{s}$ is usually called
multi-Regge, or central region for a particle $P_i$. But this
region  does not require separate consideration, because
amplitudes for production of $P_i$ in any of fragmentation regions
are applicable in it. Actually, natural bounds for domains of
applicability of these amplitudes are $\alpha_i\ll 1$ for the
gluon and $\beta_i\ll 1$ for the quark fragmentation regions.
Therefore, it is sufficient to consider two regions:
$1\geq\beta_{i}\geq \sqrt{|k^2_{i{\perp}}|/s}$ and
$1\geq\alpha_i\geq \sqrt{ |k^2_{i{\perp}}|/s}$.  For a brevity, we
will say that in the first (second) case  there are two particles
in the gluon (quark) fragmentation region. Moreover, the symmetry
with respect to $\alpha\leftrightarrow \beta$ in the definition of
the regions permits to consider only one of them. Indeed,  as
regards the inverse reaction ($Q'+G'\rightarrow G+Q$) the names of
the regions must be changed; therefore  their contribution to the
discontinuity  are related by the substitutions $Q\leftrightarrow
Q',\;G\leftrightarrow G'$ and complex conjugation. We will
consider the gluon fragmentation region.

In this region amplitudes in the right hand side of the unitarity
relation (\ref{Disc})  can be  written, in accordance with our
agreement about denotations,   as
\begin{equation}\label{multi g}
\Ac^{P_1P_2P_3}_{AB}
=\frac{2s}{q^2_{1}}\Gamma^r_{\{P_1P_2\}A}\Gamma^r_{P_3B} \,
\end{equation}
and
\begin{equation}\label{multi q}
\Ac^{P_1P_2P_3}_{AB} =\Gamma_{\{P_1P_2\}A}\frac{-\hat q_{2}}{q^2_{2}}
\Gamma_{P_3B} \,
\end{equation}
for  gluon and quark exchanges respectively, with the same
vertices $\Gamma^r_{P_3B}$ and $\Gamma_{P_3B}$ as for the elastic
amplitudes, but taken now in Born approximation only. Therefore,
as well as in the two-particle contribution, only $t$-channel
colour triplet  does survive in positive signature, since
\begin{equation}\label{project}
\mathrm{P}^{(+)}\sum\limits_{P_3}
\Gamma^r_{P_3\,Q}\left(\Gamma_{P_3\,G'}\right)^*
=-\mathrm{P}^{(+)}\sum\limits_{P_3}\Gamma_{P_3\,Q}\left(\Gamma^r_{P_3\,G'}\right)^*=
-\frac{g^2}{2} t^rt^{G'}\hat e^*_{G'}u(p_Q)\,.
\end{equation}
The vertices $\Gamma^r_{\{P_1P_2\}A}$ and $\Gamma_{\{P_1P_2\}A}$
can be found  in Refs. \cite{F2003} and \cite{LV} respectively.
Actually, they can be easily calculated, since are given by Born
amplitudes  for processes $A+R\rightarrow P_1+P_2$, where $R$ is
either a gluon (for  $\Gamma^r_{\{P_1P_2\}A}$) with momentum
$q_1$, colour index $r$ and polarization vector $p_2/s$ ($p_1/s$)
for a particle $A$ with predominant momentum components along
$p_1$ ($p_2$), or a quark (for $\Gamma_{\{P_1P_2\}A}$) with
momentum $q_2$ and omitted quark wave function. An important point
is that corresponding light-cone gauge (see (\ref{axial2})) must
be taken not only for real gluons, but for virtual ones as well.
Note that the hermitian property of Born amplitudes gives the
relations
\begin{equation}\label{hermitian}
\left(\Gamma^r_{\{P_1P_2\}Q'}\right)^*=\Gamma^r_{Q'\{P_1P_2\}}\,\;\;
\left(\Gamma_{\{P_1P_2\}Q'}\right)^\dagger=\Gamma_{Q'\{P_1P_2\}}\,\gamma^0\,,
\end{equation}
which we  use in the following.

Let us denote
\begin{equation}
\sum_{P_1P_2}\left(\Gamma^r_{\{P_1P_2\}G}\Gamma_{Q'\{P_1P_2\}}-
\Gamma_{\{P_1P_2\}G}\Gamma^r_{Q'\{P_1P_2\}}\right)t^r= g^4\bar
u(p_{Q'})t^G F_G~.  \label{FG}
\end{equation}
The particles $P_1$ and $P_2$ can be two gluons ($G_1G_2$), quark
and gluon ($Q_1G_2$), and quark and antiquark ($Q_1\bar Q_2$).
Evidently, only the first (second) term in the left hand side of
(\ref{FG}) contributes in the first and third  cases (in the
second case). Using that the phase space element $d\Phi_3$ in the
gluon fragmentation region looks like
\begin{equation}
\label{phase_3}
 d\Phi_3=\delta(1-\beta_1-\beta_2)\frac{d\beta_1 d\beta_2}{4s\beta_1\beta_2}
\frac{d^{D-2}k_1d^{D-2}k_2}{(2\pi)^{2D-3}}\,,
\end{equation}
with the help of (\ref{multi g})--(\ref{project}) we obtain the
contribution of this region to $\Delta_s$ (\ref{Disc}) in the form
\begin{multline}\label{Delta G}
\Delta_s^{(3)G}=-\frac{\epsilon^2(-q^2)^{1-2\epsilon}}{\pi^{(D-2)}\Gamma_{\epsilon}^2}
\int\!\! \frac{d\beta_1 d\beta_2}{\beta_1\beta_2}\,\frac{d^{D-2}k_1
d^{D-2}k_2}{q^2_{1}q^2_{2}}F_G\hat q_{2}\hat
e^*_{G'}\\
\mbox{}\times\delta(1-\beta_1-\beta_2)\theta(\beta_1-
\sqrt{\frac{|k^2_{1\perp}|}{s}})\theta(\beta_2-
\sqrt{\frac{|k^2_{2\perp}|}{s}})\,, %
\end{multline}
where $q_1$ and $q_2$ are the momenta of $t$-channel gluon and
quark respectively, $q_1+q_2 =q$.  As it was already pointed out,
total three--particle contribution can be written then as
\begin{equation}\label{Delta 3 total}
\Delta_s^{(3)}=\Delta_s^{(3)G}+ \bar \Delta_s^{(3)G}
(G\rightarrow G')\,, %
\end{equation}
where $\bar \Delta = \gamma^0\Delta^\dagger \gamma^0$.  Note that
logarithms of $s$ appear in $\Delta_s^{(3)G}$ from integration
over $\beta_i$ of those terms in $F_G$ which don't turn into zero
at $\beta_i \rightarrow 0$. It is always possible to rewrite $F_G$
as a sum of terms which turn into zero either at $\beta_1=0$, or
at $\beta_2=0$. For the first (second) of them the limitation
$\beta_1> \sqrt{|k^2_{1{\perp}}|/{s}}$ ($\beta_2>
\sqrt{|k^2_{2{\perp}}|/{s}}$) can be taken away; at that the
change of variables $k_{1{\perp}}= \beta_1 l_{1{\perp}}$
($k_{2{\perp}}= \beta_2 l_{2{\perp}}$) turn out often to be
useful, and we meet integrals
\begin{equation}\label{log_fac}
\int\limits^1_{\sqrt{|k^2_{i{\perp}}|/s}}\frac{d\beta_i}{\beta_i}(1-\beta_i)^\delta
=\frac{1}{2}\ln\frac{s}{|{k}^2_{i{\perp}}|}+\psi(1)-\psi(\delta+1)\,,
\end{equation}
where  $\delta$ is proportional to $\epsilon$.

Calculation of integrals without $\ln{|{k}^2_{i{\perp}}|}$ does
not represent a problem. The list of  basis integrals is presented
in Appendix \textbf{B}. Contrary, two of integrals with
$\ln{|{k}^2_{i{\perp}}|}$ can not be expressed in terms of
elementary functions at arbitrary $\epsilon$. Besides of the
integral $I_0$ (see (\ref{integral0})) encountered already in the
calculations of the two-loop gluon Regge trajectory,  here we meet
a new nontrivial integral $I_1$ (see (\ref{integral1})) which is
considered in Appendix \textbf{C}.

In the following we will use Eqs. (\ref{FG}), (\ref{Delta G}),
(\ref{Delta 3 total})  and the denotations:
\begin{equation}\label{vectors}
k_{iG}=(k_i-\beta_i p_G)_{\perp}\,,\;k_{iQ'}=(k_i-\beta_i
p_{Q'})_{\perp}\,,\; k_{12}=-k_{21}=(\beta_2k_1-\beta_1
k_2)_{\perp}\,.
\end{equation}
It's easy to understand that $k_{iA}$ ($i=1,2$) is  transverse
part of $k_i$  with respect to $p_A,\,p_2$ plane, multiplied by
$\beta_A$.

Since the integration in (\ref{Delta G}) is symmetric with respect
to exchange  $k_1\leftrightarrow k_2$, we will systematically omit
in $F_G$ contributions antisymmetric relative to this exchange,
without further  reminding.
\subsection{Fragmentation into two gluons}
The vertex for production of the gluons $G_1,G_2$ with colour
indices $i_1, \;i_2$ by the initial gluon $G$ can be written as
\begin{equation}\label{G-GG}
\Gamma^r_{\{G_1G_2\}G} = g^2
\left\{\left(T^{G}T^{r}\right)_{i_1i_2}\left[\gamma^{\mu\nu}
(k_{1G})-\gamma^{\mu\nu}(k_{12})\right]e^{\ast}_{1\mu}e^\ast_{2\nu}
+(1\leftrightarrow2) \right\}\,,\nonumber\\
\end{equation}
where
\begin{equation}
\gamma^{\mu\nu}(p)=\frac{2}{p^2_{}}\left[\beta_1\beta_2
g^{\mu\nu}_{{\perp}}(e_G p)_{}- \beta_1e^{\mu}_{G}p^{\nu}_{} -
\beta_2p^{\mu}_{} e^{\nu}_{G}\right]\, . \label{gammaggg}
\end{equation}
The vertex for a quark exchange can be taken in \cite{LV}. Using
Sudakov parametrization and omitting  terms with rightmost $\hat
p_2$ we have for it
\begin{equation} \label{Q'-GG}
\Gamma_{Q'\{G_1G_2\}} = - g^2\bar
u(p_{Q'})\left\{t^{i_1}t^{i_2}\left[\gamma_{12}^{\mu\nu}-
\gamma_{[12]}^{\mu\nu}\right]e_{1\mu}e_{2\nu} +(1\leftrightarrow2)
\right\}\,,
\end{equation}
where
\[
\gamma_{12}^{\mu\nu}=\frac{1}{k_{1Q'}^{2}}\left(\beta_1\hat k_{1Q'}
\gamma^{\mu}_{{\perp}}-2k_{1Q'}^{\mu}\right)\gamma^{\nu}_{{\perp}},
\]
\begin{equation}
\gamma_{[12]}^{\mu\nu} =\frac{2}{k_{12}^2}\left[\beta_1\beta_2
g^{\mu\nu}_{{\perp}}\hat k_{12}-
\beta_1\gamma^{\mu}_{{\perp}}k_{12}^{\nu} -
\beta_2k_{12}^{\mu}\gamma^{\nu}_{{\perp}}\right]\, .
\label{gammaqgg}
\end{equation}
Note that lower indices of $\gamma^{\mu\nu}$-vertices in
(\ref{Q'-GG}) are determined by sequences of colour matrices in
corresponding group factors, and square brackets in
$\gamma_{[12]}^{\mu\nu}$  emphasize its antisymmetry with respect
to the permutation $1\leftrightarrow 2$, as well as its relation
with the colour factor $[t^{i_1}t^{i_2}]$.

As was pointed already, only the first term in the left side part
of (\ref{FG}) does contribute in the case of two-gluon production.
Putting there the vertices (\ref{G-GG}) and (\ref{Q'-GG}) and
performing summation over spin and colour of intermediate gluons,
after simple colour algebra we obtain for the two-gluon
contribution to $F_G$
\begin{equation}\label{FGG}
F^{GG}_G=-\frac{N_c^2}{4}
\left[\left(\gamma_{\mu\nu}(k_{1G})-\gamma_{\mu\nu}(k_{12})\right)
\left(\gamma_{12}^{\mu\nu}- \gamma_{[12]}^{\mu\nu}\right)
+(1\leftrightarrow2) \right]\,.%
\end{equation}
One should pay attention that all terms in (\ref{FGG}) correspond
to planar diagrams. It is an important property of a colour
triplet state in the $t$-channel which strongly simplifies
calculations.

The convolutions entering in (\ref{FGG}) give
\begin{multline}
\gamma_{\mu\nu}(p)\gamma_{12}^{\mu\nu}
=\frac{2}{k^{2}_{1Q'}p^2}\{%
\hat{e}_{G}[2\beta_2{(k_{1Q'}p)}+\beta_1(1-2\beta_2)\hat k_{1Q'}
\hat{p}] +4\beta_1\beta_2{(e_G k_{1Q'})}\hat{p}\\
\mbox{}+\beta_1\beta_2[\beta_1(D-2)-4]{(e_G p)}
\hat k_{1Q'}\}\,;%
\end{multline}
\begin{multline}
\gamma_{\mu\nu}(p)\gamma_{[12]}^{\mu\nu} =\frac{-4}{k^{2}_{12}p^2} \{
\beta_1\beta_2(2-\beta_1\beta_2(D-2)){(e_G p)}\hat k_{12}\\
\mbox{}-2\beta_1\beta_2{(e_G k_{12})}\hat{p}
-(1-2\beta_1\beta_2){(k_{12}p)}\hat{e}_G\}\,.%
\end{multline}
We obtain the two-gluon contribution in $\Delta_s^{(3)G}$ by
substituting (\ref{FGG}) in (\ref{Delta G}) and performing
integration.  Note that if we integrate over all phase space in
(\ref{Delta G}), we have to take into account equivalence of
produced gluons by the factor $1/2!$. With account of the quark
fragmentation region according to (\ref{Delta 3 total}) we obtain,
denoting the contributions related to the terms
$\gamma_{\mu\nu}(p)\gamma^{\mu\nu}_{N}$ in $F_G^{QQ}$ (\ref{FGG})
as $\Delta_{N\ast p}\;$:
\begin{align}
\Delta_{12\ast
k_{1G}}&=\frac{N_c^2}{4}\left\{2\left[\frac{2}{\epsilon}-\frac{3}{1+2\epsilon}+
\Psi_2+\Ln+\frac{5}{2\epsilon}\Xr+\frac{1}{2}(I_0+I_1)\right]\Qc+\frac{2\epsilon}{1+2\epsilon}\,\Ec\right\}\,;\nonumber\\
\Delta_{12\ast
k_{12}}&=\frac{N_c^2}{4}\Xr\left\{2\left[\frac{3}{1+2\epsilon}-\frac{3}{\epsilon}
-\Ln-\Psi\right]\Qc-\frac{2\epsilon}{1+2\epsilon}\,\Ec\right\}\,; \quad
\Delta_{[12]\ast
k_{12}}=0\,;\label{Deltas GG}\\
\Delta_{[12]\ast
k_{1G}}&=-\frac{N_c^2}{4}\Xr\left\{2\left[\Ln+\Psi%
+\frac{5}{2\epsilon}-\frac{4}{1+2\epsilon}+\frac{1}{(1-\epsilon)(1+2\epsilon)(3+2\epsilon)}\right]\Qc\right.\nonumber\\
\mbox{}&\phantom{\frac{4\epsilon^2}{(3+2\epsilon)(1-\epsilon)(8+2\epsilon)}}
-\left.\frac{4\epsilon^2}{(3+2\epsilon)(1-\epsilon)(1+2\epsilon)}\,\Ec\right\}\,,\nonumber
\end{align}
where
\begin{equation} \label{integral0}
 I_0=-\frac{\epsilon^2(-q^2)^{1-2\epsilon}}{\Gamma^2_{\epsilon}\pi^{(D-2)}}\int
 \frac{d^{D-2}k_{1}d^{D-2}k_2\;\,\;{q}^2}{{k}^2_{1\perp}
 ({k}_1-{q})_{\perp}^2
{k}^2_{2\perp}({k}_2-{q})_{\perp}^2}%
\ln\frac{{q}^2}{({k}_1-{k}_2)_{\perp}^2}\,,
\end{equation}
\begin{equation} \label{integral1}
 I_1=-\frac{\epsilon^2(-q^2)^{1-2\epsilon}}{\Gamma^2_{\epsilon}\pi^{(D-2)}}
\int
\frac{d^{D-2}k_1d^{D-2}k_2\;\;({k}_1-{k}_2)_{\perp}^2}{{k}^2_{1\perp}
({k}_1
-{q})_{\perp}^2{k}^2_{2\perp}({k}_2-{q})_{\perp}^2}%
\ln\frac{{q}^2}{({k}_1-{k}_2)_{\perp}^2}\,.
\end{equation}
Here and below we use denotations
\begin{equation}\label{psi2 and psi}
\Psi_2=2[\psi(1)-\psi(1+2\epsilon)]\,,\;\;
\Psi=\psi(1-2\epsilon)+\psi(1+3\epsilon)-\psi(1+\epsilon)-\psi(1)+\Psi_2\,.
\end{equation}
The contribution $\Delta_s^{GG}$ of fragmentation into two gluons
 to $\Delta_s^{(3)}$ is
\begin{equation}\label{Delta GG}
\Delta_s^{GG}= \Delta_{12\ast k_{1G}}+\Delta_{12\ast
k_{12}}+\Delta_{[12]\ast k_{1G}}~.
\end{equation}
\subsection{Fragmentation into quark and gluon}
Let us denote  particles produced in the gluon fragmentation
region $Q_1$ and $G_2$ and their momenta $k_1$ and $k_2$
respectively. The vertex $\Gamma_{\{Q_1\,G_1\}G}$ can be obtained
by crossing and appropriate substitutions from
$\Gamma_{Q'\{G_1G_2\}}$ (\ref{Q'-GG}). We will represent it as
\begin{equation} \label{G-QG}
\Gamma_{\{Q_1\,G_2\}G} =- g^2\bar
u(k_{1})\left\{t^{G}t^{i_2}\left[\gamma_{G2}^{\nu}-
\gamma_{[G2]}^{\nu}\right]e^*_{2\nu}
+t^{i_2}t^{G}\left[\gamma_{2G}^{\mu}+
\gamma_{[G2]}^{\mu}\right]e^*_{2\mu} \right\}\,,
\end{equation}
where $\gamma_{G2}^{\nu},\;\gamma_{[G2]}^{\nu}$ and
$\gamma_{2G}^{\nu}$ are obtained from
$\gamma_{12}^{\mu\nu},\;\gamma_{[12]}^{\mu\nu}$ and
$\gamma_{21}^{\nu\mu}$ respectively  by substitutions
\begin{equation} \label{substitutions1}
\beta_1\rightarrow \frac{1}{\beta_1}\,,\;\;\beta_2\rightarrow
-\frac{\beta_2}{\beta_1}\,,\;\;k_{1Q'}\rightarrow
-\frac{k_{1G}}{\beta_1}\,,\;\;k_{12}\rightarrow
\frac{k_{2G}}{\beta_1}\,,\;\;k_{2Q'}\rightarrow
\frac{k_{12}}{\beta_1}\,,
\end{equation}
and convolution with $e^{\mu}_G$.  The vertex
$\Gamma^r_{Q'\{Q_1G_2\}}$ can be found in \cite{F2003} and
represented as
\begin{equation}\label{QG-Q}
\Gamma^r_{Q'\{Q_1G_2\}}=g^2\bar{u}(p_{Q'})%
\frac{\hat{p}_2}{s}\left\{t^{i_2}t^r\left(L_{\mu}(k_{2Q'})+
L_{\mu}(k_{1Q'})\right)+ t^r t^{i_2} \left(
L_{\mu}(k_{12})-L_{\mu}(k_{1Q'})
 \right)\right\}e^{\mu}_2 u(k_1)\,, %
\end{equation}
where
\begin{equation}\label{bar lmu}
 L^{\mu}(p)=\frac{\beta_2\hat p_{}\gamma^{\mu}_{\perp}
-2p^{\mu}_{}}{p^2_{}}\,.  %
\end{equation}
In the case of quark-gluon production only the second term in the
right hand side  of (\ref{FG}) does contribute. Using vertices
(\ref{G-QG}) and (\ref{QG-Q}),  we obtain for the quark-gluon
contribution to $F_G$ after summation over spin and colour
\begin{equation}\label{FQG}
F^{QG}_G=-\beta_1\sum\limits_{{ij}}
L_{\mu}(k_{ij})\left[C^{ij}_{G2}
\gamma^{\mu}_{G2}+C^{ij}_{2G}\gamma^{\mu}_{2 G}%
+C^{ij}_{[G2]}\gamma^{\mu}_{[G2]} \right]\,.%
\end{equation}
where $ij$ takes on values $1Q',\;2Q', \, 12$ and
\begin{alignat}{3}\label{c1}
       &C^{1Q'}_{G 2}=\frac{1}{4}\,,\;\;& &C^{2Q'}_{G 2}
=\frac{1}{4}\left(1+\frac{1}{N_c^2}\right)\,,\;\;& %
&C^{12}_{G 2}=\frac{1}{4N_c^2}\,;\nonumber\\
    &C^{1Q'}_{2 G}=\frac{1}{4}\,,\;\; & &C^{2Q'}_{2 G}=\frac{1}{4N_c^2}\,, & %
&C^{12}_{2 G}=-\frac{1}{4}\left(1-\frac{1}{N_c^2}\right)\,;\\
    &C^{1Q'}_{[G2]}=0\,, \;\;& &C^{2Q'}_{[G2]}=-\frac{1}{4}\,, &
    &C^{12}_{[G2]}=-\frac{1}{4}\,.\nonumber
\end{alignat}
Note that the term  $C^{2Q'}_{G2} L_{\mu}(k_{2Q'})
\gamma^{\mu}_{G2}$  here corresponds to a non-planar diagram and
leads to a complicated integral. Fortunately,  it is cancelled by
respective contribution from quark-antiquark production, as we
will see in the next subsection.

For the products $ L_{\mu}(p)\gamma^{\mu}_{mn}$ we have after some
algebra:
\begin{align}\label{circ-2}
 L_{\mu}(p)\gamma^{\mu}_{G 2}&=\frac{1}{k^2_{1G} p^2}
\{2\beta_1\beta_2(D-2)
{(e_Gk_{1G})}\hat{p}-\beta_2(D-6)\hat{p}\hat{k}_{1G} \hat{e}_{G}
-2\hat{e}_{G}\hat{k}_{1G}\hat{p}\}\,;\nonumber\\
 L_{\mu}(p)\gamma^{\mu}_{2 G} &= \frac{1}{k^2_{12} p^2}
\{\beta_2(\beta_2(D-2)-4)\hat{p}\hat{k}_{12}+4{(k_{12}p)}\}
\hat{e}_G\,;\nonumber\\
 L_{\mu}(p)\gamma^{\mu}_{[G2]}&=\frac{2\beta_2}{k^2_{2G}
p^2} \left\{\left(\frac{\beta_2}{\beta_1}-1\right)\hat{e}_G\hat{p}
\hat{k}_{2G}+
(\beta_2(D-2)-4){(e_G k_{2G})}\hat{p}\right.\\
\mbox{}&+ \left.\frac{2}{\beta_2}{(k_{2G}p)}\hat{e}_G+4{(e_G p)}
\hat{k}_{2G}
\right\}\,.\nonumber%
\end{align}
The quark-gluon contribution to $\Delta_s^{(3)G}$ is given by
(\ref{Delta G}) with (\ref{FQG}) instead of $F_G$.  Denoting
$\Delta_{m\,n\cdot ij}$ the contributions proportional
$C^{ij}_{m\,n}$ in (\ref{FQG}), after  integration  we obtain for
them with account of the quark fragmentation region according to
(\ref{Delta 3 total})
\begin{align*}
&\Delta_{G 2\cdot 1Q'}=\frac{1}{4}\left\{%
2\left[\frac{3-2\epsilon}{1+2\epsilon}-\Psi_2-\Ln-\frac{5}{2\epsilon}\Xr-%
\frac{1}{2}(I_0+I_1)\right]\Qc+\frac{2\epsilon}{1+2\epsilon}\,\Ec\right\}\,;\\%
&\Delta_{G 2\cdot 12}=\frac{1}{4N_c^2}\Xr\left\{%
-2\left[\Ln+\Psi+\frac{1}{\epsilon}-\frac{3-2\epsilon}{1+2\epsilon}\right]\Qc+\frac{2\epsilon}{1+2\epsilon}\,\Ec\right\}\,;
\end{align*}
\begin{align}
&\Delta_{2 G\cdot 1Q'}=\frac{1}{4}\Xr\left\{%
2\left[\Ln+\Psi+\frac{3}{2\epsilon}-\frac{3}{2(1+2\epsilon)}\right]\right\}\Qc\,;\nonumber\\
&\Delta_{2G\cdot 2Q'}=\frac{1}{4N_c^2}\Xr\left\{%
2\left[\frac{3-2\epsilon}{2(1+2\epsilon)}-\frac{1}{\epsilon}\right]\right\}\Qc\,;\qquad%
\Delta_{2 G\cdot 12}=0\,;\label{Deltas QG}
\end{align}
\begin{align*}
&\Delta_{[G 2]\cdot 2Q'}=\frac{1}{4}\left\{%
-2\left[-\frac{3}{1+2\epsilon}+\frac{2}{\epsilon}+\Psi_2+\Ln-\frac{\Xr}{\epsilon}-\frac{1}{2}I_1\right]\Qc%
-\frac{2\epsilon}{1+2\epsilon}\,\Ec\right\}\,;\\
&\Delta_{[G 2]\cdot 12}=\frac{1}{4}\Xr\left\{%
2\left[\Ln+\Psi+\frac{3}{\epsilon}-\frac{3}{1+2\epsilon}\right]\Qc+\frac{2\epsilon}{1+2\epsilon}\,\Ec\right\}\,.
\end{align*}
The contribution $\Delta_{G\,2\cdot 2Q'}$ is not presented here,
since it is cancelled with  respective contribution from
quark-antiquark production, as was pointed out. The quark-gluon
 contribution to $\Delta_s^{(3)}$ is given by
\begin{equation}\label{Delta QG}
\Delta_s^{QG}= \sum\limits_{{ij},{mn}}\Delta_{mn\cdot
ij}\,,\;\;\;\; {ij}=1Q',\;2Q',12\,;\;\;\;\;{mn}=G2, \;2G,\;[G
2]~.
\end{equation}
\subsection{Fragmentation into quark--antiquark pair}
As well as in the two-gluon case, only the first term in
(\ref{FG}) exists. We denote momenta of particles $Q_1$ and $\bar
Q_2$ in the gluon fragmentation region  $k_1$ and $k_2$
respectively. The vertex $\Gamma^r_{\{Q_1\,\bar Q_2\}G}$ can be
obtained by crossing and corresponding substitutions from
$\Gamma^r_{Q'\{Q_1G_2\}}$ (\ref{QG-Q}):
\begin{equation} \label{G-QQ}
\Gamma^r_{\{Q_1\,\bar Q_2\}G} = g^2\bar u(k_{1})%
\frac{\hat{p}_2}{s}\left\{t^{G}t^r\left(L(k_{1G})- L(k_{12})\right)+
t^r t^{G}\left( L(k_{12})+
L(k_{2G})\right)\right\}v(k_2)\,, %
\end{equation}
where
\begin{equation}\label{bar l}
L(p)= \frac{-\hat p \,\hat e_G +2\beta_1(p\, e_G) }{p^2}\,.   %
\end{equation}

The vertex $\Gamma_{Q'\{Q_1\,\bar Q_2\}}$ can be obtained from
\cite{LV}; a direct calculation does not encounter difficulties as
well. We have
\begin{equation} \label{Q'-QQ}
\Gamma_{Q'\{Q_1\,\bar Q_2\}} = g^2 \left[
\bar{v}(k_2)t^c\gamma^\sigma u(k_1) \bar{u}(p_{Q'})t^c%
A_{\sigma}-\bar{u}(p_{Q'})t^c%
\gamma^\sigma u(k_1)\cdot \bar{v}(k_2)%
t^c B_{\sigma}%
\;\delta_{Q_1 Q}\right]\,, %
\end{equation}
where  $\delta_{Q_1 Q}$  shows that the last contribution exists
only when an intermediate quark has the same flavour as the
initial one, and  the values $A_{\sigma}$ and $B_{\sigma}$ can be
written as
\begin{equation}%
A^{\sigma}
=\frac{-\beta_1\beta_2}{k_{12}^2}\left(\gamma^\sigma_\perp-
(\hat{k}_{1}+\hat{k}_{2})_\perp\frac{2p^\sigma_2}{s}\right) \,;\;
B^{\sigma}
=\frac{\beta_1}{k^{2}_{1Q'}}\left(\gamma^{\sigma}_\perp+{(\hat{k}_1
-\hat{p}_{Q'})_\perp}%
\frac{2p^{\sigma}_2}{s\beta_2}\right)\,.%
\end{equation}
By substituting the  vertices (\ref{G-QQ}) and (\ref{Q'-QQ}) into
(\ref{FG}), after summation over spin and colour states of
intermediate particles  we obtain for the quark-antiquark
contribution to $F_G$

\begin{equation}\label{FQQ}
F^{QQ}_G=\sum\limits_{{ij}}\left[C^{ij}_{A} tr\left(\hat
k_1\frac{\hat p_2}{s} L(k_{ij})\hat
k_2\gamma_{\sigma}\right)A^{\sigma} -C^{ij}_{B}\gamma_{\sigma}\hat
k_1\frac{\hat p_2}{s}
L(k_{ij})\hat k_2 B^{\sigma} \right]\,,%
\end{equation}
where $ij$ takes on values $1G,\;2G, \, 12$ and
\begin{alignat}{3}\label{c2}
    C^{1G}_{A}&=\frac{N_c}{8}n_f\,, \;\;\;& C^{2G}_{A}&=-\frac{N_c}{8}n_f\,,
    \;\;\; \;\;\;\;\;\;\;\;& C^{12}_{A}&=-\frac{N_c}{4}n_f\,;\\
    C^{1G}_{B}&=\frac{1}{4N_c^2}\,, & C^{2G}_{B}&=\frac{1}{4}\left(1
    +\frac{1}{N_c^2}\right), \;\;\;&
    C^{12}_{B}&=\frac{1}{4}\,.\nonumber
\end{alignat}
Taking into account, as usually, external Dirac spinors, we obtain
\[
 tr\left(\hat k_1\frac{\hat p_2}{s} L(p)\hat
k_2\gamma_{\sigma}\right)A^{\sigma}
=\frac{2\beta_1\beta_2}{p^2 k_{12}^2}\left(%
{(k_{12}p)}\hat{e}_G-{(e_G
k_{12})}\hat{p}+(1-4\beta_1\beta_2){(e_G
p)}\hat{k}_{12}\right)\,;\\
\]
\begin{equation}\label{AB}
\gamma_{\sigma}\hat k_1 \frac{\hat p_2}{s}L(p)\hat k_2
B^{\sigma}=\frac{-\beta_1}{k^{2}_{1Q'} p^2}
\{2\beta_1\beta_2(D-2){(e_G p)}\hat{k}_{1Q'}%
-\beta_2(D-6)\hat{k}_{1Q'}\hat{p}\hat{e}_G%
-2\hat{e}_G\hat{p}\hat{k}_{1Q'}\}\,.
\end{equation}
The quark-antiquark contribution to $\Delta_s^{(3)G}$ is given by
(\ref{Delta G}) with (\ref{FQQ}) instead of $F_G$. As have been
mentioned above, in $F_G^{QQ}$  we also have a term corresponding
to a non-planar diagram.  In (\ref{FQQ}) it stands with the
coefficient $C^{2G}_{B}$. Note, that $C^{2G}_{B}=C^{2Q'}_{G2}$
(see (\ref{c1})). Moreover, comparing the first equations in
(\ref{circ-2}) and the second in (\ref{AB}), one can see that  the
substitution
\begin{equation}\label{subst}
k_{1Q'}\rightarrow -k_{2Q'}\,,\quad k_{2G}\rightarrow -k_{1G}\,
\end{equation}
turns $\gamma_{\sigma}\hat k_1 (\hat p_2/s)L(k_{2G})\hat k_2
B^{\sigma}$ into  $\beta_1 L_{\mu}(k_{2Q'}) \gamma^{\mu}_{G2}$
with opposite sign. Note that for the quark-antiquark contribution
 $q_2$ in (\ref{Delta G}) is equal
$p_{Q'\perp}-k_{1\perp}-k_{2\perp}=\beta_2 q -k_{1Q'}-k_{2G}$. At
the substitution (\ref{subst}) it turns into $\beta_2 q
+k_{2Q'}+k_{1G}=k_{1\perp}+k_{2\perp}-p_{G\perp}$, that is just
the  $t$-channel quark momentum for the quark-gluon contribution.
An important point is that the theta-functions in (\ref{Delta G})
can be omitted for the contributions of the nonplanar diagrams due
to convergence of integrals, after  that the substitution
(\ref{subst}) does not influence on the integration region.
Therefore these contributions cancel each other.

Performing integration and denoting $\Delta_{A\cdot ij}$ and
$\Delta_{B\cdot ij}$  the contributions to (\ref{FQQ})
proportional $C^{1j}_{A}$ and $C^{1j}_{B}$ respectively, we obtain
with account of the quark fragmentation region according to
(\ref{Delta 3 total}):%
\[ 
\Delta_{A\cdot 1G}= \frac{N_c}{8}n_f\Xr\left\{%
\frac{-2}{1+2\epsilon}\left(1-\frac{1}{(1-\epsilon^2)(3+2\epsilon)}\right)\Qc%
-\frac{4\epsilon^2}{(1-\epsilon^2)(1+2\epsilon)(3+2\epsilon)}\,\Ec\right\}\,;%
\]%
\[%
\Delta_{A\cdot 2G}=\Delta_{A\cdot 1G}\,;\;\;\;\Delta_{A\cdot
12}=0\,;
\]%
\[%
\Delta_{B\cdot 1G}=\frac{1}{4N_c^2}\left\{%
2\left[-\frac{3-2\epsilon}{1+2\epsilon}+\Psi_2+\Ln-\frac{\Xr}{\epsilon}-\frac{1}{2}I_1\right]\Qc%
-\frac{2\epsilon}{1+2\epsilon}\,\Ec\right\}\,;
\]%
\begin{equation}\label{Deltas Q barQ}
\Delta_{B\cdot 12}=\frac{1}{4}\Xr\left\{%
-2\left[\frac{3-2\epsilon}{1+2\epsilon}-\Ln-\Psi-\frac{1}{\epsilon}\right]\Qc-\frac{2\epsilon}{1+2\epsilon}
\,\Ec\right\}\,.
\end{equation}
Therefore, the quark-antiquark  contribution to $\Delta_s^{(3)}$
can be written as
\begin{equation}\label{Delta QQ}
\Delta_s^{QQ}= 2\Delta_{A\cdot 1G}+\Delta_{B\cdot 1G}
+\Delta_{B\cdot 12}+\Delta_{B\cdot 2G}\,,
\end{equation}
where $\Delta_{B\cdot 2G}=-\Delta_{G 2\cdot 2 Q'}$, as it was
shown above. Because of cancellation of these contributions in
\begin{equation}\label{Delta 3}
\Delta_s^{(3)}= \Delta_s^{GG}+\Delta_s^{GQ}+\Delta_s^{QQ}\,
\end{equation}
$\Delta_{B\cdot 2G}$  is not presented in  (\ref{Deltas Q barQ}),
as well as $\Delta_{G 2\cdot 2 Q'}$ in (\ref{Deltas QG}).
\section{Two--loop correction to the quark trajectory}
The total discontinuity $\Delta_s$ is given by the sum of the
contributions of two- and three-particle intermediate states in
the unitarity relation. The two-particle contribution
$\Delta^{(2)}_s$ is given explicitly by (\ref{Delta 2}). All
necessary contributions to $\Delta^{(3)}_s$ (\ref{Delta 3}) are
calculated  in the preceding section. They are given by
(\ref{Delta GG})and (\ref{Deltas GG}) for fragmentation into two
gluons, by (\ref{Delta QG}) and (\ref{Deltas QG}) for
fragmentation into quark and gluon, and  by (\ref{Delta QQ}),
(\ref{Deltas Q barQ}) for fragmentation into  quark and antiquark.
Now we can compare the calculated discontinuity with the form
(\ref{Delta}) required by the quark Reggezation.   First of all we
note coincidence of the terms with $\ln s$. Actually this
coincidence must be expected, since it is required by the quark
Reggeization in LLA, which was already checked on this level. Much
more important  is that   the calculated discontinuity has the
same  helicity structure as $\Delta_R$ (\ref{Delta}), with the
same coefficient at  the structure $\Ec$. It is a serious argument
in favour of validity of the Reggeization hypothesis in NLA. Then
comparing coefficients at the structure $\Qc$ we obtain
\begin{multline}\label{total_tr}
\dt{2}=C_F\left\{-n_f\frac{\Xr(1+\epsilon)}{(1+2\epsilon)(3+2\epsilon)}
+N_c\left(\frac{1}{2}I_0+I_1-\Xr[\psi(1+\epsilon)-\psi(1+2\epsilon)]\right.\right.\\
\mbox{}+\left.\left.\frac{7\Xr}{2\epsilon}+ \frac{\Xr(11+7\epsilon)}{2(1+2\epsilon)(3+2\epsilon)}\right)
+2C_F\left(-\frac{1}{2}I_1+\psi(1)-\psi(1-\epsilon)\right.\right.\\
\mbox{}+\left.\left.(2-\Xr)[\psi(1+\epsilon)-\psi(1+2\epsilon)]-
\frac{1+\Xr}{\epsilon}-\frac{(1-\Xr)(3-\epsilon)}{2(1+2\epsilon)}\right)\right\}\,,
\end{multline}
where $\Xr$ is given by (\ref{Xr}), $I_0$ and $I_1$ are defined in
(\ref{integral0}) and (\ref{integral1}).

The two-loop corrections to the quark Regge trajectory
$j_Q=1/2+\delta_{\mathrm{T}}(t)$ is determined by Eqs.
(\ref{g-traj}), (\ref{total_tr}) at arbitrary space-time dimension
$D=4+2\epsilon$. Unfortunately, at arbitrary $D$ the integrals
$I_0$ and $I_1$ can not be written in terms of elementary
functions.  In the limit $\epsilon\rightarrow 0$ we have for them
(see Appendix \textbf{C})
\begin{equation}
 I_0=\frac{1}{\epsilon}+15\psi^{(2)}(1)\,\epsilon^2+\mathcal{O}(\epsilon^3)\,,\;\;
 I_1=-\frac{4}{\epsilon}+6\psi^{(2)}(1)\,\epsilon^2+\mathcal{O}(\epsilon^3)\,,
\end{equation}
where $\psi^{(2)}$ means the second derivative of the
$\psi$-function. With this result and with the proportion
$\psi^{(2)}(1)=-2 \zeta_3$, where $\zeta_n$ is the Riemann
Zeta-function, from Eq.\eqref{total_tr}   we obtain for the
two-loop correction up to non-vanishing at $\epsilon\rightarrow 0$
terms
\begin{equation}\label{delta_exp}
\dt{2}=C_F\left\{\beta_0-K\epsilon+\left(\left(\frac{202}{27}-9\,\zeta_3\right)N_c-\frac{28}{27}n_f%
+8\,\zeta_3\,C_F\right)\epsilon^2\right\}+\mathcal{O}(\epsilon^3)\,,
\end{equation}
where%
\[
\beta_0=\frac{11}{6}N_c-\frac{2}{3}n_f\,,\quad
K=\left(\frac{67}{18}-\zeta_2\right)N_c-\frac{5}{9}n_f \,, %
\]
in agreement with the result of Ref. \cite{BD-DFG}.
\section{Summary}
In this paper we have checked  compatibility of the quark
Reggeization  hypothesis with the $s$-channel unitarity by the
explicit two-loop calculations and have found  in the case of
massless quarks the two--loop correction to the quark trajectory
at arbitrary space--time dimension $D=4+2\epsilon$.  The
$\epsilon$--expansion of the correction gives the result obtained in
\cite{BD-DFG}. We have calculated the two-loop $s$--channel
discontinuity of the backward quark-gluon scattering amplitude
with positive signature keeping non-logarithmic terms and have
proved that  only a colour triplet part of the amplitude does
survive at NLO as well as at LO.  It was shown that the calculated
discontinuity has a form required by the Reggeization hypothesis.
The  two--loop correction to the trajectory has been obtained from
comparison of the calculated discontinuity with the  Reggeized
form. In the case of massive quarks the trajectory can be found by
the same method, since all necessary quantities  for such
calculation are known.

The cancellation of contributions of colour states different from
triplet in positive signature is not restricted by  the two-loop
approximation, but is a general property of NLA. Therefore in this
approximation, as well as in LLA,  real parts of amplitudes with
positive signature are completely determined by Reggeized quark
contributions.

Note that testing of the quark Reggeization  performed up to now
is rather limited.  Even in LLA self-consistency  of the
hypothesis was shown only in a particular case of elastic
quark-gluon scattering, although in all orders of $\alpha_s$.  In
NLA it is tested in the same process only in order $\alpha_s^2$. A
possible way of more strict testing and, in principle, a complete
proof  is examination  of "bootstrap" conditions on the Reggeized
quark vertices and the trajectory in QCD. These conditions appear
from comparison of Reggeized form of discontinuities of amplitudes
with quark exchanges with the discontinuities calculated with use
of the $s$-channel unitarity.

We have used dimensional regularization for both infrared and
ultraviolet divergences and the bare coupling constant
$g=g_s\mu^{\epsilon}$, so that besides of infrared poles in
$\epsilon$ our result contains the ultraviolet poles. To remove
them it is sufficient to express the bare coupling through
renormalized one. In the $\overline{\mathrm{MS}}$ renormalization
scheme
\begin{equation}
g =  g_\mu\mu^{-\epsilon}
\left\{1+\beta_0\frac{g^2_{\mu}\Gamma(1-\epsilon)}{\epsilon
{(4\pi)}^{2+\epsilon}}\right\}\,,
\end{equation}
where $g_{\mu}$ is the renormalized  coupling  at the
normalization point $\mu$.
\begin{acknowledgments}
A.V.B. grateful to \fbox{M.I. Kotsky}
for helpful discussions. V.S.F. thanks the
Alexander von Humboldt foundation for the research award, the
Universit\"at Hamburg and DESY, the Dipartimento di Fisica
dell'Universit\`a della Calabria and the Istituto Nazionale di
Fisica Nucleare - gruppo collegato di Cosenza for their warm
hospitality while a part of this work was done. This work is
supported in part by INTAS and in part by the Russian Fund of
Basic Researches.
\end{acknowledgments}
\appendix
\section{}
\noindent For convenience of a reader  we list here the integrals
encountered at the calculation of $\Delta_s^{(2)}$. Note that
everywhere below we use Euclidean transverse momenta and omit the
transversality sign.
\begin{align}
J^i_1&=\int\!\!\frac{d^{D-2}k\; k^i}{(k^2)^{1-\epsilon}(k-q)^2}=
q^i\;\frac{\pi^{(D-2)/2}}{\epsilon\,(q^2)^{1-2\epsilon}}\,\Gamma_\epsilon\cdot\Xr\,;\\
J^i_2&=\int\!\!\frac{d^{D-2}k\; k^i}{(k^2)^{1-\epsilon}(k-q)^2}
\ln\frac{s}{k^2}=J^i_1\left[\Ln+\Psi_1\right]\,;
\end{align}
\begin{align}
J^i_3&=\int\!\!\frac{d^{D-2}k\; q^i}{(k^2)^{1-\epsilon}(k-q)^2}=%
\frac{3}{2}J^i_1\,;\\
J^i_4&=\int\!\!\frac{d^{D-2}k\;
q^i}{(k^2)^{1-\epsilon}(k-q)^2}\ln\frac{s}{k^2} =
\frac{3}{2}J^i_2+\frac{1}{4\epsilon}J^i_1\,;
\end{align}
\begin{equation}
J^\mu_5=\int\!\!\frac{d^{D-2}k\; k^l k^m q^i}{(k^2)^{2-\epsilon}(k-q)^2} =
\frac{1}{\epsilon(1-\epsilon)}\left(\frac{q^k q^l}{q^2}(1-2\epsilon) +
\frac{\delta^{k\,l}}{4}\right)J^i_1\,.
\end{equation}
Remind that
\begin{gather}
\Gamma_{\epsilon}=\frac{\Gamma^2(1+\epsilon)\Gamma(1-\epsilon)}{\Gamma(1+2\epsilon)}\,; \quad %
\Xr=\frac{\Gamma(1-2\epsilon)\Gamma^2(1+2\epsilon)}{\Gamma(1+\epsilon)\Gamma(1+3\epsilon)\Gamma^2(1-\epsilon)}\,;\nonumber\\
\Psi_1=\psi(1+3\epsilon)+\psi(1-2\epsilon)-\psi(1+2\epsilon)-\psi(1-\epsilon)\,.\label{gam}
\end{gather}
\section{}
Apart from $I_0$  (\ref{integral0}) and $I_1$ (\ref{integral1}),
considered in the next Appendix, integrals required for
calculation of $\Delta_s^{(3)}$ can be transformed to ones listed
below.  Using the denotations (\ref{gam}), (\ref{psi2 and psi})
and
\begin{equation}\label{lam}
\Lambda=\frac{\pi^{(D-2)}}{\epsilon^2({q}^2)^{1-2\epsilon}}\,;
\end{equation}
we have
\begin{equation}
K_1=\int\!\!\frac{dk^{(D-2)}dp^{(D-2)}}{\Lambda(\Gamma_\epsilon)^2}\frac{\hat{e}_G
(k q) \hat{p}}
{k^2(k-q)^2 p^2(p-q)^2} = \hat{e}_G \hat{q}\,;
\end{equation}%
\begin{align}
K_2&=\int\!\!\frac{dk^{(D-2)}dp^{(D-2)}}{\Lambda(\Gamma_\epsilon)^2}\frac{(e_G
k) (\hat{k}-\hat{q}) \hat{p}} {k^2(k-q)^2 p^2(p-q)^2}\\
\mbox{}&=-\hat{e}_G\hat{q}\frac{1}{2(1+2\epsilon)} + {(e_G q)}
\frac{\epsilon}{1+2\epsilon}\,;
\end{align}
\begin{align}
K_3&=\int\!\!\frac{dk^{(D-2)}dp^{(D-2)}}{\Lambda(\Gamma_\epsilon)^2}\frac{\hat{e}_G
\hat{k} (\hat{k}-\hat{q}) \hat{p}} {k^2(k-q)^2 p^2(p-q)^2}
\ln\frac{s}{(k-p)^2}\nonumber\\
\mbox{}&=\hat{e}_G\hat{q}\left[\frac{1}{2}(I_0+I_1)+\Xr\frac{5}{2\epsilon} + \Ln \right]\,;
\end{align}
\begin{equation}
K_4=\int\!\!\frac{dk^{(D-2)}dp^{(D-2)}}{\Lambda(\Gamma_\epsilon)^2}\frac{\hat{e}_G
(\hat{k}-\hat{q}) \hat{k} \hat{p}} {k^2(k-q)^2p^2(p-q)^2}
\ln\frac{s}{(k-p)^2} = \hat{e}_G\hat{q}\left[-\frac{1}{2}I_1-\frac{\Xr}{\epsilon} +
\Ln \right]\,;
\end{equation}
\begin{align}
K_5&=\int\!\!\frac{dk^{(D-2)}dp^{(D-2)}}{\Lambda(\Gamma_\epsilon)^2}\frac{(e_G
k) \hat{k}\hat{p}} {k^2(k-p)^2p^2(p-q)^2} = -\left(\hat{e}_G
\hat{q}\frac{1}{2} + (e_Gq)(1+\epsilon)\right)\frac{\Xr}{(1+2\epsilon)}\,;\\
K_6&=\int\!\!\frac{dk^{(D-2)}dp^{(D-2)}}{\Lambda(\Gamma_\epsilon)^2}\frac{(e_G
k) (\hat{k}-\hat{p}) (\hat{p}-\hat{q})} {k^2(k-p)^2 p^2(p-q)^2} =
\left(\hat{e}_G\hat{q}\frac{1}{4} + (e_G
q)\epsilon^2\right)\frac{\Xr}{(1-\epsilon)(1+2\epsilon)}\,;
\end{align}
\begin{equation}
K_7=\int\!\!\frac{dk^{(D-2)}dp^{(D-2)}}{\Lambda(\Gamma_\epsilon)^2}\frac{\hat{e}_G
\hat{k} p^2}
{k^2(k-p)^2p^2(p-q)^2} = \hat{e}_G\hat{q}\;\Xr\,;
\end{equation}
\begin{align}
K_8&=\int\!\!\frac{dk^{(D-2)}dp^{(D-2)}}{\Lambda(\Gamma_\epsilon)^2}\frac{\hat{e}_G
\hat{k} p^2} {k^2(k-p)^2 p^2(p-q)^2} \ln\frac{s}{(k-p)^2}\nonumber\\
\mbox{}&=\hat{e}_G \hat{q} \left[\Ln+\Psi-\Psi_2+ \frac{1}{\epsilon} \right]\Xr\,;
\end{align}
\begin{equation}
K_9=\int\!\!\frac{dk^{(D-2)}dp^{(D-2)}}{\Lambda(\Gamma_\epsilon)^2}\frac{\hat{e}_G
(kp) (\hat{p}-\hat{q})} {k^2 (k-p)^2 p^2 (p-q)^2} =
-\frac{1}{2}\hat{e}_G\hat{q}\;\Xr\,;
\end{equation}
\begin{align}
K_{10}&=\int\!\!\frac{dk^{(D-2)}dp^{(D-2)}}{\Lambda(\Gamma_\epsilon)^2}\frac{\hat{e}_G
(kp) (\hat{p}-\hat{q})} {k^2 (k-p)^2 p^2 (p-q)^2}
\ln\frac{s}{(k-p)^2}\nonumber\\
\mbox{}&= -\frac{1}{2}\hat{e}_G\hat{q}\left[\Ln+
\Psi-\Psi_2 + \frac{3}{2\epsilon}\right]\Xr\,.
\end{align}
\section{}
At arbitrary $D\neq4$ the integrals $I_0$  (\ref{integral0}) and
$I_1$ (\ref{integral1}) can be expressed only through infinite
series. They belong to the class of integrals which was studied
particulary in \cite{BGK}. The first of them has already appeared
in the calculation of the two-loop correction to the gluon Regge
trajectory \cite{FFK96}, where its limit at $\epsilon\rightarrow
0$ was found:
\begin{equation}\label{I_0}
I_0=\int\frac{d^{D-2}k_1d^{D-2}k_2}{\Lambda(\Gamma_\epsilon)^2}%
\frac{q^2}{k^2_1(k_1-q)^2 k^2_2(k_2-q)^2}%
\ln\frac{q^2}{(k_1-k_2)^2}=%
\frac{1}{\epsilon}+15\psi^{(2)}(1)\,\epsilon^2+\mathcal{O}(\epsilon^3)\,.
\end{equation}
Here $\Gamma_\epsilon$ and  $\Lambda$ are given in \eqref{gam} and
\eqref{lam}; $\psi^{(2)}$ means the second derivative of $\psi$.
We have obtained  the limit of the second integral:
\begin{equation}\label{I_1}
I_1=\int \frac{d^{D-2}k_1d^{D-2}k_2}{\Lambda(\Gamma_\epsilon)^2}%
\frac{(k_1-k_2)^2}{k^2_1(k_1-q)^2k^2_2(k_2-q)^2}%
\ln\frac{q^2}{(k_1-k_2)^2}=%
-\frac{4}{\epsilon}+6\psi^{(2)}(1)\,\epsilon^2+\mathcal{O}(\epsilon^3)\,.
\end{equation}
Below some details of the calculation are given.

Representing the integral as
\begin{equation}
 I_1=\left.\frac{d}{d\nu}I(\nu)\right|_{\nu=0}\,,\;\;
{I}(\nu)=\int \frac{d^{D-2}k_1d^{D-2}k_2}{\Lambda(\Gamma_\epsilon)^2}%
\frac{(q^2)^\nu}{k^2_1(k_1-q)^2
k^2_2(k_2-q)^2[(k_1-k_2)^2]^{\nu-1}}\,,
\end{equation}
we have from  Eqs. (1), (4), (9) of \cite{BGK}:
\begin{equation}\label{SS}
I(\nu)=\frac{2\epsilon^2}{2\epsilon-1}\frac{\nu^{-1}(\nu-\epsilon)}{(\Gamma_\epsilon)^2} G_2(1,1-\epsilon+\nu)\,%
G_2(1,\nu)\cdot\left(\nu(\nu-1)\lim\limits_{b\rightarrow0}
S(\epsilon-1,b,2\epsilon-\nu,\nu-1-\epsilon)\right)\,,
\end{equation}
where
\begin{equation}
G_2(\alpha_1,\alpha_2)=G_1(\alpha_1)G_1(\alpha_2)G_1(2+2\epsilon
-\alpha_1-\alpha_2)\,,\;\;
G_1(\alpha)=\frac{\Gamma(1+\epsilon-\alpha)}{\Gamma(\alpha)}\,.
\end{equation}
The function S(a,b,c,d) is defined by Eqs. (17), (16) and  (10) of
\cite{BGK}:
\begin{equation}
S(a,b,c,d)=\frac{\pi\cot(\pi c)}{H(a,b,c,d)}-\frac{1}{c}-\frac{b+c}{bc}%
F(a+c,-b, -c,b+d))\,,
\end{equation}
where
\begin{equation}
H(a,b,c,d)=\frac{\Gamma(1+a)\Gamma(1+b)\Gamma(1+c)\Gamma(1+d)\Gamma(1+a+b+c+d)}%
{\Gamma(1+a+c)\Gamma(1+a+d)\Gamma(1+b+c)\Gamma(1+b+d)}\,,
\end{equation}
and $F(a,b,c,d)$ is expressed through the  generalized
hypergeometric function ${_3F_2}$:
\begin{equation}
F(a,b,c,d)={_3F_2}\left[\begin{array}{c}%
-a,-b,1\\
1+c,1+d \end{array};1\right] -1\,.
\end{equation}
The limit $b\rightarrow 0$ in (\ref{SS}) can be easily taken. We
can use also for this purpose Eq.(12) of \cite{BGK} and obtain
with our  values of parameters
\begin{equation}
\nu(\nu-1)\lim\limits_{b\rightarrow0} S(\epsilon-1,b,2\epsilon-\nu,\nu-1-\epsilon)=
S_1(\nu)+S_2(\nu)\,,
\end{equation}
where
\begin{equation}\label{S1}
S_1(\nu)=\pi\cot[\pi(2\epsilon-\nu)]\frac{\Gamma(3\epsilon-\nu)\Gamma(1+\nu)}{\Gamma(\epsilon)\Gamma(2\epsilon-1)}%
-\frac{\nu(\nu-1)}{2\epsilon-\nu}\,;
\end{equation}
\begin{equation}\label{S2}
S_2(\nu)=\nu(\nu-1)\frac{3\epsilon-1-\nu}{(\epsilon-1)(\nu-2\epsilon)}F(1-\epsilon,2\epsilon -\nu,\nu
-2\epsilon, \nu-2)\,.
\end{equation}
With these denotations we have
\begin{equation}\label{int}
I_1=\frac{d}{d\nu}\left(S_0(\nu)(S_1(\nu)+S_2(\nu))
\right)|_{\nu=0}\,,
\end{equation}
where
\begin{equation}\label{S0}
S_0(\nu)=\frac{\epsilon^2}{(\Gamma_\epsilon)^2}\frac{2}{2\epsilon-1}\frac{\Gamma^2(\epsilon)\Gamma(2\epsilon-\nu)\Gamma(1-2\epsilon+\nu)%
\Gamma(1-\nu+\epsilon)}{\Gamma(3\epsilon-\nu)\Gamma(1+\nu)\Gamma(1+2\epsilon-\nu)}\,.
\end{equation}%
We need to know $S_i(0)$ and $S'_i(0)\,,\;\;\;i=0\div 2$. For
$i=0,1$ they are easily obtained from (\ref{S0}), (\ref{S1}):
\begin{align}
S_{0}(0)&=\frac{3}{2\epsilon-1}\Xr\,,\quad%
S'_{0}(0)=S_{0}(0)\left(\frac{1}{6\epsilon}+\Psi-\Psi_2\right)\,;\nonumber\\
S_{1}(0)&=2\pi\cot[2\pi\epsilon]\frac{\epsilon(2\epsilon-1)}{3}%
\frac{\Gamma(1+3\epsilon)}{\Gamma(1+\epsilon)\Gamma(1+2\epsilon)}\,;\label{S_01_dif}\\
S'_{1}(0)&=S_{1}(0)\left(\psi(1)-\psi(1+3\epsilon)+\frac{1}{3\epsilon}
+\frac{2\pi}{\sin(4\pi\epsilon)}\right)+\frac{1}{2\epsilon}\,, \nonumber
\end{align}
where $\Xr$ is defined in (\ref{gam}), $\Psi$ and $\Psi_2$  in
(\ref{psi2 and psi}). To find $S_2(0)$ and $S'_2(0)$ we use the
integral representation
\begin{equation}
F(a,b,c,d)=-1+\frac{\Gamma(1+d)\Gamma(1+c)}{\Gamma(d)\Gamma(-b)
\Gamma(1+c+b)}\int\limits^1_0 dx \,(1-x)^{d-1}\int\limits^1_0dz\,
z^{-b-1}(1-z)^{c+b}(1-zx)^{a}\,
\end{equation}
which follows from the standard representation for the
hypergeometric functions.  Performing  integration over $x$ by
parts three times we obtain
\begin{multline}
S_2(\nu)=(3\epsilon-1-\nu)\left(\frac{\nu}{1-2\epsilon+\nu}+\frac{\epsilon}{2-2\epsilon+\nu}\right.\\
\mbox{}+\left.\epsilon\int\limits^1_0\!\!\int\limits^1_0dx\,dz\,z^{1-2\epsilon+\nu}(1-x)^\nu%
\frac{d}{dx}(1-zx)^{-1-\epsilon}\right)\,,
\end{multline}
so that
\begin{gather}\label{S_2_dif}
S_{2}(0)=(1-2\epsilon)\frac{\Gamma(1-\epsilon)\Gamma(1-2\epsilon)}{\Gamma(1-3\epsilon)}\,,\\
S'_{2}(0)=\frac{S_{2}(0)}{1-3\epsilon}+(3\epsilon-1)\left(\frac{1}{1-2\epsilon}-
\frac{\epsilon}{4(1-\epsilon)^2}+\epsilon(J_1+J_2)\right)\,,
\end{gather}
where%
\begin{align}
J_1&=\int\limits_0^1\!\!\!\int\limits^1_0dx\,dz\,z^{1-2\epsilon} \ln
z\frac{d}{dx}(1-zx)^{-1-\epsilon} \,, \nonumber \\
J_2&=\int\limits_0^1\!\!\!\int\limits^1_0 dz\,dx\,z^{1-2\epsilon} \ln(1-x)
\frac{d}{dx} (1-zx)^{-1-\epsilon} \,.\label{J12}
\end{align}
The integral $J_1$ can be easily found:
\begin{equation}\label{J1}
J_1=\frac{1}{4(1-\epsilon)^2}+\frac{\Gamma(1-\epsilon)
\Gamma(2-2\epsilon)}{\Gamma(2-3\epsilon)}\frac{\left(\psi(2-3\epsilon)-
\psi(2-2\epsilon)\right)}{\epsilon}\,.
\end{equation}
For the integral $J_2$, replacing $d/dx$ by $(z/x)d/dz$ in  the
representation (\ref{J12}) and integrating over $z$ by parts we
obtain in the limit $\epsilon \rightarrow 0$
\begin{equation}\label{J2}
J_2\simeq \frac{-1}{\epsilon^2}+1+\psi^{(1)}(1)+\epsilon\left(
5-\psi^{(1)}(1)+\frac{\psi^{(2)}(1)}{2}\right)\,.
\end{equation}
Using in (\ref{int}) the results (\ref{J1}), (\ref{J2}),
\eqref{S_01_dif} and \eqref{S_2_dif} we get the final result
(\ref{I_1}).

\newpage
\ \

\begin{figure}[t!]
\setcaptionmargin{0mm}
\onelinecaptionsfalse
\centering
\setlength{\unitlength}{1mm}
 \begin{picture}(50,40)
   \includegraphics[width=50mm]{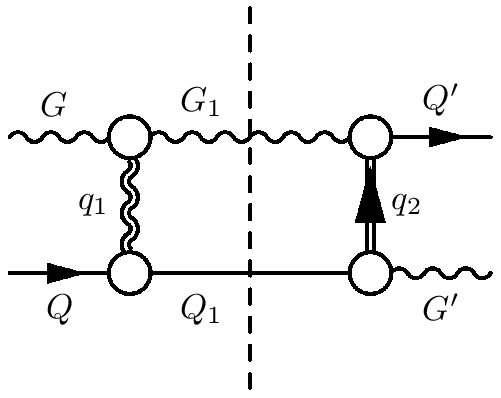}
   \put(-35,1){a)}
 \end{picture}
\hspace{2.2cm}
 \begin{picture}(50,0)
   \includegraphics[width=50mm]{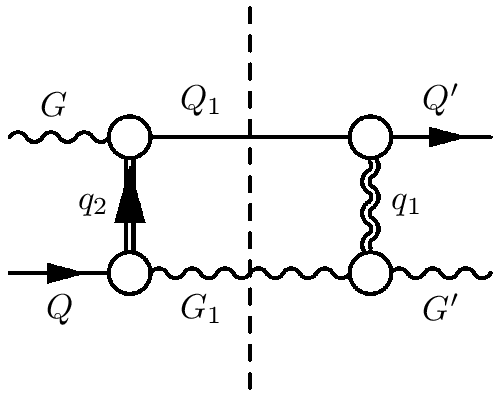}
   \put(-35,1){b)}
 \end{picture}
\captionstyle{flushleft}\caption{Schematic representation of the two--particle
contribution to the $s$--channel discontinuity of the backward
quark--gluon scattering amplitude.  The doubled lines represent
Reggeized quark and gluon, and the blobs --- PPR
vertices.}\label{pic_2}
\end{figure}
\ \

\begin{figure}[h]
\setcaptionmargin{0mm}
\onelinecaptionsfalse
\centering
\setlength{\unitlength}{1mm}
 \begin{picture}(50,40)
   \includegraphics[width=50mm]{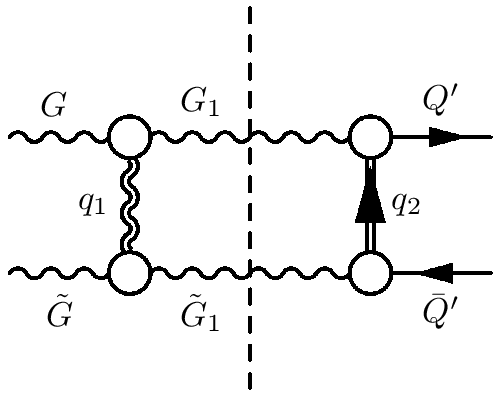}
   \put(-35,1){a)}
 \end{picture}
\hspace{2.2cm}
 \begin{picture}(50,0)
   \includegraphics[width=50mm]{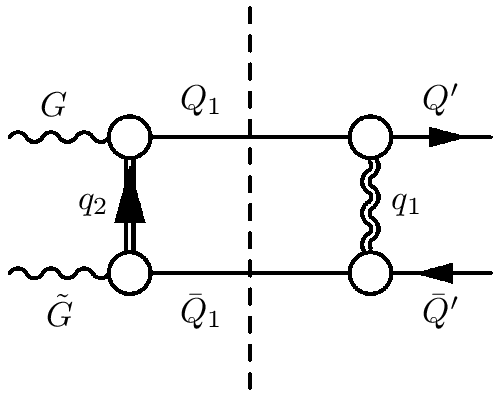}
   \put(-35,1){b)}
 \end{picture}
\captionstyle{normal}\caption{Schematic representation of the two--particle
contribution to the cross channel discontinuity of the backward
quark--gluon scattering amplitude.}\label{pic_3}
\end{figure}

\end{document}